\documentclass{aa}

\usepackage{comment}
\usepackage{graphicx} 
\usepackage[normalem]{ulem} 
\usepackage[varg]{txfonts}
\usepackage{amssymb}
\usepackage{amsmath}
\usepackage{mathrsfs}
\usepackage{enumerate}
\usepackage{subfigure}
\usepackage{multirow}
\usepackage{siunitx}
\usepackage[colorlinks=true,allcolors=blue]{hyperref}
\usepackage[dvipsnames]{xcolor}
\usepackage{siunitx}

\newcommand{\Figss}[2]{Figures~\ref{fig:#1} -- \ref{fig:#2}} 
\newcommand{\cf}{\textit{cf.} }
\newcommand{\eg}{\textit{e.g.} }
\newcommand{\ie}{\textit{i.e.} }

\newcommand{\llnr}[1]{{\bf \color{magenta}{[]}} \color{black}}

\defcitealias{Touresse24}{T24}

\title{Dynamics of a switchback-like magnetic deflection induced by a propagating coronal jet}
\titlerunning{Switchback-like magnetic deflection induced by coronal jet}
\authorrunning{Touresse et al.}

 \author{
 J. Touresse \inst{1}\orcid{0009-0003-5268-5128},
 E. Pariat\inst{2, 1}\orcid{0000-0002-2900-0608}, 
 C. Froment \inst{3}\orcid{0000-0001-5315-2890},
 P. F. Wyper\inst{4}\orcid{0000-0002-6442-7818}
 }

 \institute{Sorbonne Universit\'e, \'Ecole polytechnique, Institut Polytechnique de Paris, Universit\'e Paris Saclay, Observatoire de Paris-PSL, CNRS, Laboratoire de Physique des Plasmas (LPP), 75005 Paris, France \\
         \and French-Spanish Laboratory for Astrophysics in Canarias (FSLAC), CNRS IRL 2009, Instituto de Astrofísica de Canarias, 38205, La Laguna, Tenerife, Spain \\
		 \and LPC2E, OSUC, Univ Orleans, CNRS, CNES, F-45071 Orleans, France\\
		 \and Durham University, Department of Mathematical Sciences, Stockton Road, Durham, DH1 3LE, UK
      }

\date{Received 19 February 2026 / Accepted 27 July 2026}

\abstract
{The Parker Solar Probe (PSP) mission revealed the ubiquity of switchbacks (SBs), \ie, sharp and localised magnetic deflections, in the nascent solar wind. This finding has sparked significant interest in their origin and formation mechanisms. Among the proposed scenarios, a solar origin associated with solar jets has emerged. Solar jets are collimated, impulsive and sharp-edged features associated with magnetic reconnection and generate an Alfvénic torsional wavefront propagating into the heliosphere.}
{In this study, we investigate whether a realistic spacecraft sampling of a propagating solar jet can produce synthetic in-situ signatures consistent with PSP-observed switchbacks. We also characterise the three-dimensional (3D) magnetic topology of the underlying structure and assess how the magnitude of the magnetic deflection evolves with the surrounding environment.}
{We perform a 3D magnetohydrodynamic simulation of a solar jet propagating into the ambient solar wind. A virtual spacecraft trajectory, constructed from the actual PSP orbit during its 22$^{\text{nd}}$ perihelion, is embedded within the simulation to generate synthetic in-situ measurements.}
{The simulation reproduces SB-like magnetic signatures with durations and profiles consistent with PSP observations, confirming earlier results obtained using simplified trajectories. For the first time, we directly visualise the fully 3D magnetic topology of a jet-driven SB region. Two distinct families of magnetic field lines are identified: one exhibiting a single magnetic deflection and another showing two sequential deflections propagating along the same field line. In both configurations, the deflections act as outward-propagating Alfvénic structures. The magnitude of the magnetic deflection is strongly modulated by the background plasma $\beta$, with significant amplification in low-$\beta$ regions and a more moderate evolution where $\beta \sim 1$.}
{These results demonstrate that propagating solar jets can naturally produce Alfvénic, SB-like structures when sampled along realistic in-situ trajectories. The magnetic topology of SB regions is inherently 3D and structurally diverse. Their evolution depends on the surrounding plasma $\beta$, with complete magnetic reversals requiring longer residence times in low-$\beta$ environments. This study supports the interpretation that at least a fraction of the SBs observed by PSP may have a solar origin and provides new constraints for interpreting PSP measurements in terms of their underlying magnetic structure.}


\begin{document}

\maketitle
\nolinenumbers

\section{Introduction} \label{sec:Introduction}

Intermittent events characterised by a sharp deflection of the magnetic field vector away from the Parker spiral and an associated increase in the solar wind (SW) radial velocity are known as switchbacks \citep[SBs; ][]{Bale19, Kasper19}. 
Earlier missions such as Helios, Ulysses, and ACE  seldom detected them at or beyond Earth orbit \citep[][]{Yamauchi04, Landi06, Gosling09, Matteini14}. However, Parker Solar Probe \citep[PSP;][]{Fox16} and subsequently Solar Orbiter \citep[SolO;][]{Muller20, Garcia21} provided unprecedented observations demonstrating their ubiquity \citep[][]{Bale19, Kasper19, DudokdeWit20, Horbury20, Fedorov21, Suen23, HouC24, Badman26} in the inner heliosphere. This surge of interest mainly stems from their likely key role as markers of coronal heating and solar wind acceleration mechanisms. 

SBs are Alfvénic variations corresponding to pronounced deflections of the magnetic field from its pristine solar wind direction, while the total magnetic field intensity remains approximately conserved \citep[$|B| \sim \mathrm{const}$; \eg][]{Horbury20, Mozer20, Laker21, Larosa21}. SBs are typically associated with enhancements in the radial proton bulk velocity, often referred to as "spikes", also called plasma jets \citep[][]{Bale19, Kasper19}. Most SBs correspond to deflections of less than $90 ^\circ$ \citep[\eg][]{DudokdeWit20, Bandyopadhyay22}, whereas a small portion involves complete polarity reversals of the radial magnetic field, hereafter referred to as full-reversal SBs.  In contrast to previous studies \citep[e.g.][]{Bale19, Kasper19, Akhavan24} that restricted the term "switchbacks" to Alfvénic inversions, \ie where the radial field $B_r$ changes sign, following \eg \citet[][]{DudokdeWit20, Bizien23, Touresse24} we adopt a broader definition of SBs taking into account this continuum of deflections. This choice is motivated by observational evidence that there is no typical angle, and SBs exhibit similar statistical and morphological properties across a wide range of magnitudes of the deflection \citep[\eg][]{DudokdeWit20, Bizien23, Badman26}. For instance, the thickness and nature of their boundaries do not appear to depend strongly on the magnitude of the magnetic deflection \citep[][]{Bizien23}. Full magnetic reversals may therefore be viewed as the high-magnitude tail of a broader population of Alfvénic magnetic deflections rather than as a distinct class of events.

SBs exhibit not only a wide range of deflection angles but also a broad diversity in their temporal and spatial characteristics. Their duration spans from a few seconds to several hours \citep[][]{DudokdeWit20, Pecora22, Badman26}. They frequently appear in clusters lasting hours to days \citep[\eg][]{Shi22, Pecora22}, separated by quieter magnetic-field intervals \citep[][]{Bale19, Kasper19, Horbury20}. However, as \citet{Laker21} emphasises, the apparent duration of an individual SB is strongly influenced by the spacecraft trajectory through the structure, introducing an observational bias. This is consistent with the spatial structure of a SB assessed by \citet{Horbury20}, which describes SBs as elongated or tube-like structures oriented predominantly along the radial direction. \citet{Horbury20} estimates that they are tens of thousands of kilometres across, with high aspect ratios, consistent with a long and thin morphology. \citet{Laker21} extended these results to multiple encounters, showing that the observed periods correspond to structures approximately $50\,000$ km wide, with aspect ratios of order ten, and with their major axis slightly deflected from the radial direction along the Parker spiral. \citet{Mozer21} further shows that SBs expand more rapidly in the radial direction than in the tangential one, increasing their aspect ratios. Taken together, these studies indicate that SBs are generally aligned with the Parker spiral and could propagate along it.

Despite the wealth of in-situ data, the spacecraft samples the solar wind at a single location, making it difficult to separate the spatial structure of a SB from its temporal evolution. Simulations can complement these observations by providing a complete three-dimensional perspective, helping to corroborate their geometry while disentangling temporal and spatial variations.

The widespread observations of SBs in PSP measurements have spurred extensive interest in their origin and formation mechanisms, giving rise to several theories that can broadly be classified into two groups \citep[][]{Wyper26}. 
The first category encompasses "in-situ" formation mechanisms, in which SBs arise through processes occurring within the solar wind. These include interactions between SW streams \citep[][]{Ruffolo20, Schwadron21}, the interaction between velocity shears and pre-existing Alfvénic waves \citep[][]{Toth23} or the development of Alfvénic wave modes driven by SW expansion \citep[][]{Squire20, Mallet21, Shoda21, Squire22, Johnston22, Bowen25}.
The second category, sometimes referred to as "ex-situ" formation mechanisms, posits that SBs originate from mechanisms happening in the low solar atmosphere. Among those low atmosphere mechanisms, interchange reconnection, either occurring gradually in a steady-state coronal background \citep[][]{Fisk20, Zank20, Liang21} or taking place in an impulsive manner, as with coronal jets or jetlets events \citep[][]{Sterling20, Raouafi23, Bizien25}. 

Solar jets are impulsive, collimated features that are observed to propagate outward throughout the solar atmosphere, and are detected across a wide range of wavelengths, from $H_{\alpha}$ to soft X-rays, spanning temperatures between $10^4$ and $10^7$ K. They are considered as mid-scale events in solar activity, and their spatial scales span from about a few to $10^5$ km. Depending on the observing wavelength, their size, and their locations, these phenomena have been labelled as spicules, surges, chromospheric jets, jetlets, coronal jets, EUV jets, or X-ray jets \citep[see reviews of ][]{Raouafi16, ShenY21}.

Recent studies indicate that full-reversal SBs are less frequent in the sub-Alfvénic wind \citep[\eg][]{Pecora22, Bandyopadhyay22, Akhavan24}. This supports the idea of an in-situ origin, or at least the involvement of in-situ processes in their formation. Nevertheless, the possibility that solar jets could act as sources for SBs has gained considerable attention. This renewed interest stems from the ubiquity and intermittent occurrence of jet-like events throughout the solar atmosphere   \citep[\eg][]{Raouafi23, Kumar23,  HouC24, Bizien25}. A crucial question, however, concerns whether these jets, or their byproducts, can propagate sufficiently far into the heliosphere to generate signatures consistent with the ones of SBs. Whereas jets launched in closed-field environment or in the vicinity of active regions are confined and unlikely to escape, those in coronal holes may readily reach the inner heliosphere, driving Alfvénic waves generated by impulsive magnetic reconnection along open magnetic field lines. Observationally, some large-scale jets have been seen to reach the upper corona \citep[][]{Patsourakos08, Nistico09, Pucci13, Moore15}, as well as some intermediate-size EUV jets observed in white-light during the 2018 eclipse \citep[][]{Hanaoka18}. The observation from the Metis coronograph in \citet{Telloni22} may also be the signature of a particularly large jet event propagating upward. Yet, such coronographic observations have recorded jets only sporadically. The results in \citet[][]{Hanaoka18} study tend to suggest that our ability to detect jet propagation in coronal images - whether through direct imaging or coronographs - is strongly limited by instrument sensitivity. The fact that smaller jets are not observed by imaging instruments does not necessarily imply that these events fail to propagate far into the solar wind. Hitherto, \citet{Parenti21} report the only deterministic one-to-one correspondence between a given coronal jet and in-situ measurements. This underscores the intrinsic challenges of performing such a direct comparison, even for large jets, given the uncertainties in terms of connectivity \citep[see also][]{Bizien25}. 
Because of observational and magnetic field mapping limitations, the link between jet-like activity and SBs can, at present, be mainly investigated statistically. Several studies have compared the occurrence rates and properties of SBs observed during PSP perihelia with those of small-scale transients and jets in the low solar atmosphere \citep[][]{DePablos22, Raouafi23, Kumar23, HouC24, Bizien25}.
 While these comparisons indicate suggestive similarities in scale, frequency, and occurrence patterns, a definitive correspondence has yet to be demonstrated.
Another approach to investigating the link between solar jets and SBs involves numerical simulations that follow the evolution of jets from their origin to the inner heliosphere. The modelling of jet-like events is a well-established field that has been extensively studied over several decades \citep[\cf reviews of ][]{Raouafi16, ShenY21, Skirvin22, Tziotziou23}. These studies have provided major insights into jet triggering processes, particle acceleration, multi-thermal and multi-flows dynamics and comparison with observed coronal features \citep[ \eg ][for recent models]{Pariat16,Wyper19,GonzalezAviles21,Pallister21,ChenF22,ZhuJ23}. 

A number of studies have specifically focused on the propagation of jet-like events into an extended domain \citep[\eg][]{Lionello16, Karpen17, Szente17, Uritsky17, Roberts18, Touresse24}. Especially, \citet{Roberts18} analyses the jet from \citet{Karpen17} and produces synthetic in-situ measurements of magnetic and velocity fields. Their results show that the untwisting wave front corresponds to a local magnetic deflection, characterised by a decrease in the radial magnetic field component, $|B_r|$, while the total magnetic field magnitude, $|B|$, remains approximately constant. A simultaneous enhancement of the radial velocity is observed \citep[see Fig. 9 in][]{Roberts18}. Although limited to a single jet event in a strongly magnetically dominated corona ($\beta \ll 1$ throughout the domain), their study suggests that some SBs may originate from magnetic untwisting waves along solar jet-like events. 

Building on this, \citet{Touresse24} (hereafter \citetalias{Touresse24}) investigates the dynamics of the propagation of coronal jets into the solar wind using adaptive three-dimensional (3D) magnetohydrodynamic (MHD) simulations with various atmospheric conditions. \citetalias{Touresse24} confirms one of the results of \citet[][]{Pariat16}, demonstrating that self-consistent jets can indeed be generated and propagate in different low coronal plasma $\beta$ environments. Like other simulations \citep[\eg][]{Lionello16, Karpen17, Uritsky17, Roberts18}, the \citetalias{Touresse24} jets can easily propagate over tens of solar radii away from the solar surface.  As shown in \citet[][]{Karpen17, Uritsky17, Roberts18} and \citetalias{Touresse24}, the jet structure consists of a leading Alfvénic torsional wave followed by a trailing dense jet. In \citetalias{Touresse24}, magnetic reconnection is immediately followed by the formation of low-lying U-shaped magnetic field lines near the solar surface; however, these structures do not persist at higher coronal altitudes. In contrast, the leading Alfvénic torsional wave propagates outward and exhibits properties typical of SBs, although the associated deflection angle remains below  $90^\circ$.

A limitation of \citetalias{Touresse24} is that the modelled torsional wavefront was intersected by an infinitely fast virtual spacecraft. Therefore, the realism and accuracy of direct comparisons with in-situ measurements remained limited. To address this limitation, future work must incorporate more realistic spacecraft trajectories, as suggested by \citet{Lynch21, Lynch25}. The objective of the present study is to build upon \citetalias{Touresse24} by introducing realistic PSP trajectories into one of the 3D MHD simulations of \citetalias{Touresse24}. This approach enables a direct characterisation of the temporal and spatial evolution of the magnetic deflections associated with the Alfvénic wavefront of the jet.

In particular, the present study aims to address the following questions: 
\begin{itemize}
    \item Do the results of \citetalias{Touresse24} hold when a proper PSP trajectory is used: do we still obtain typical SB profiles with a consistent time scale?
    \item What is the magnetic geometry of a SB event? 
    \item How does the magnetic deflection progress radially?
    \item How does the magnitude of the deflection evolve in time/radius?
\end{itemize}

For clarity, hereafter, the term "switchback" refers to the in-situ measurements of the phenomena, either observed or sampled by the virtual spacecraft. The expression "switchback event/region" corresponds to the underlying physical structure, while "deflection" denotes the geometrical bending of the magnetic field line within this structure. In the simulations, the SB structure is generated by an untwisting jet/ a propagating wavefront/ an Alfvénic wavefront or a combination of those phenomena. The manuscript is organised as follows. Section~\ref{sec:Data} summarises the main elements of the simulation setup previously used in \citetalias{Touresse24}, along with the PSP trajectory implementation and the method employed to track magnetic deflections within SB events. Section~\ref{sec:mag_geo_sb} examines the magnetic geometry of a representative SB event. Section~\ref{sec:evol_def} describes the temporal and radial evolution of the deflections. Section~\ref{sec:intensity_def} explores how the magnitude of the deflections varies and whether it correlates with local plasma conditions. Finally, Section~\ref{sec:conclusion} summarises our findings and discusses their implications in the broader context of current SB studies.

\section{Data and methods} \label{sec:Data}

This section starts by outlining the 3D MHD simulation. We summarise the simulation setup, including both its physical and numerical aspects, as well as the main steps of its evolution (Section \ref{sec:simulation_desc}). Next, we describe the trajectory used to generate synthetic in-situ measurements, which is an adaptation of the actual PSP path during its perihelion on 24 December 2024 (Encounter 22) (Section \ref{sec:psp_traj}). Finally, we present the method to track the evolution of magnetic field lines within the torsional wave front (Section \ref{sec:track_def}), which exhibits signatures consistent with those of SBs, providing the framework for the analysis performed in Sections \ref{sec:evol_def} and \ref{sec:intensity_def}.

\subsection{Simulation model description} \label{sec:simulation_desc}

The present study extends the H$\beta$ simulation developed by \citetalias{Touresse24}. Its key features are briefly summarised below.

The simulation domain is a spherical wedge spanning [1 $\rm{R}_{\sun}$, 15 $\rm{R}_{\sun}$]$\times$[81 $\degr$, 99 $\degr$]$\times$[- 9 $\degr$, 9 $\degr$] in radius, colatitude \footnote{the angle complementary to latitude; it equals $0^{\circ}$ at the North pole and $90^{\circ}$ at the equator} and longitude with $\rm{R}_{\odot} = 7 \times 10^5$ km, the solar radius. The grid is stretched exponentially in the radial direction and spaced linearly in angle to ensure nearly cubic cells throughout the domain. The inner and outer radial boundaries are open, while the side boundaries are closed and reflective. Additionally, a line-tying boundary condition is imposed at the inner radial boundary to drive the system.

The computational grid is initialised with a base resolution of 27 × 3 × 3 blocks, each containing 8 × 8 × 8 cells. Up to four levels of mesh refinement are allowed, and the grid is initialised with higher refinement levels concentrated near the solar surface and the jet propagation region at the domain centre (see Figure 1 of \citetalias{Touresse24}).

The 3D numerical simulation is performed using the Adaptively Refined MHD Solver \citep[ARMS,][]{DeVore91,DeVore08}, which solves time-dependent, isothermal, compressible, ideal MHD equations. The ARMS code uses the PARallel Adaptive MESH refinement (PARAMESH) toolkit \citep[][]{MacNeice00}, which dynamically refines the computational grid in regions where the electric current density is locally strong. The flux-corrected transport (FCT) algorithm \citep[][]{DeVore91} is used to advance the following set of MHD equations in time on the adaptive grid solved in spherical coordinates:
\begin{gather}
    \frac{\partial \rho }{\partial t} + \nabla . (\rho \mathbf{v}) = 0, \label{eq:continuity_eq} \\
    \frac{\partial \rho \mathbf{v}}{\partial t} + \nabla . (\rho \mathbf{v v}) = - \nabla P + \rho \mathbf{g} + \frac{1}{4 \pi} (\nabla \wedge \mathbf{B} \wedge \mathbf{B}), \label{eq:momentum_eq} \\
    \frac{\partial \mathbf{B}}{\partial t } - \nabla \wedge (\mathbf{v} \wedge \mathbf{B})  = 0\label{eq:induction_eq}
\end{gather}

where $t$ is the time, $\mathbf{v}$ is the velocity field, $\rho$ is the mass density, $P$ is the thermal pressure, $\mathbf{g}$ is the gravitational acceleration, and $\mathbf{B}$ is the magnetic field. 
The energy equation is not included, as we assume an isothermal plasma. This assumption is the simplest and most computationally efficient way to generate a supersonic solar wind \citep[][]{Parker58}. To close the system, we assume the plasma follows the ideal gas law for fully ionised hydrogen.\\

The base pressure and temperature of the atmosphere are set to: $P_b =  5.00 \times 10^{-8}$ bar and $T_b = 2 \times 10^6$ K, respectively. These values are consistent with estimates derived from Yohkoh/SXT observations of coronal holes \citep[][]{Hara96, Mogilevsky97}. The corresponding base density is determined from these parameters through the ideal gas law, while the density profile in the rest of the domain is obtained by imposing a constant mass flux. The atmosphere is initialised following the steady, spherically symmetric, isothermal SW solution of \citet{Parker58}:
\begin{equation}
    \frac{v_r^2}{c_{\mathrm{s}}^2} \exp \left( 1- \frac{v_r^2}{c_{\mathrm{s}}^2} \right) = \frac{r_{\mathrm{s}}^4}{r^4} \exp \left(4-4 \frac{r_{\mathrm{s}}}{r} \right)
    \label{eq:solar_wind_radial_velocity}
\end{equation}
where the radial velocity $v_{r}$ satisfies this transonic Parker wind equation. The isothermal sound velocity is given by $c_{\mathrm{s}} = \sqrt{2 k_{\mathrm{B}} T_{\mathrm{b}} / m_{\mathrm{p}}}$, and the sonic point is located at $r_{\mathrm{s}} = G M_{\odot} m_{\mathrm{p}} / (4 k_{\mathrm{B}} T_{\mathrm{b}})$. The combined action of solar gravity and the Parker wind solution naturally establishes a radially stratified, isothermal atmosphere with a supersonic outflow consistent with coronal-hole conditions.

In addition to these hydrodynamic parameters, we impose a magnetic field configuration.
The background coronal-hole field is modelled as a Sun-centred monopole with strength $B_m = -1.5$ G at the surface. To mimic a localised parasitic polarity—typically the source of jet activity—we introduce a point dipole at a depth $d = 1 \times 10^2$ km below the surface, outside the computational domain. The dipole is aligned radially, with a maximum surface field strength of $B_d = 35$ G.
These elements produce an anemone-like shape, characterised by a dome-shaped separatix surface containing a 3D null point with its associated fan surface and two spine lines. It aligns closely with observations of jet-like events \citep[see][]{Nistico09, Liu16, Joshi20}.

Based on these parameter choices, the simulation yields a sonic surface at  $r_s = 2.9 \,  \rm{R}_{\odot}$, an Alfvén surface at $r_A = 4.6 \, \rm{R}_{\odot}$,  a plasma $\beta = 1$ surface at $r_{\beta = 1} = 4.8 \,  \rm{R}_{\odot}$.  The sonic surface marks the transition where the outflow speed of the plasma exceeds the local sound speed; it becomes supersonic. The Alfvén surface is where the local flow velocity of the plasma surpasses the local Alfvén speed, the solar wind becomes super Alfvénic. Finally, the plasma $\beta = 1$ surface represents the location where the thermal pressure equals the magnetic pressure, distinguishing the region of magnetically dominated dynamics from the thermally dominated one. 

Since the simulation is not performed in a rotating frame, corresponding to the solar rotation, the Parker spiral is not included and the background magnetic field remains purely radial. Deflection angles are therefore measured relative to the radial direction. Within the heliocentric range covered by the simulation ($1-15 R_{\odot}$), corresponding to an outer boundary located approximately $10 R_{\odot}$ beyond the plasma beta = 1 surface, the deviation of the Parker spiral from the radial direction remains only a few degrees. Therefore, the omission of the Parker spiral is not expected to significantly affect the results presented in this study.
The interaction with the solar wind produces an initial imbalance between magnetic and kinetic energies within the system. To obtain a physically consistent starting point, we conduct a relaxation phase in the absence of both dynamic grid refinement and photospheric driving. During this period, the system evolves toward a quasi-steady state characterised by near-constant magnetic and kinetic energies. The relaxation phase extends over $65\,100$ s, providing a stable equilibrium configuration before further driving is introduced.

Once the system has relaxed, we initiate the energy injection phase, where magnetic energy is supplied to the closed-field system that fuels the jet. We apply a quasi-steady rotational flow across the entire parasitic polarity at the bottom boundary. This provides a simplified representation of the  slow shearing and rotational motions observed in active regions on the Sun.

As the energy accumulation phase concludes, the system undergoes a rapid transition into an eruptive regime. This change is driven by the development of an instability and the onset of fast magnetic reconnection, which converts a fraction of the stored magnetic energy into plasma kinetic energy. This energy release naturally leads to the formation of a coronal jet. Importantly, the twist present in the system is not imposed as an external control parameter but instead emerges self-consistently through the system’s dynamical evolution \citep[][]{Pariat09a}.

The jet is produced through interchange magnetic reconnection between closed and open magnetic field lines. This process generates nonlinear Alfvénic disturbances that propagate away from the reconnection region while reconnection-driven plasma outflows develop concurrently. Together, these wave motions and plasma flows form the jet structure. As the jet propagates into the outer corona and the inner heliosphere, the leading Alfvénic perturbation compresses the ambient plasma, producing a density enhancement in the trailing jet material relative to the local solar wind background (see \citetalias{Touresse24}). Propagating at typical coronal Alfvén speeds, which can reach approximately $420$ km s$^{-1}$, the jet front traverses one solar radius in roughly $1\,300$ s.

\subsection{PSP trajectory data} \label{sec:psp_traj}

In \citetalias{Touresse24}, the simulated virtual spacecraft is considered to cross the torsional wavefront at an infinitely fast speed. Although PSP moves faster than the solar wind near perihelion, the infinite-speed assumption is an over-simplification. 
In most of the encounters, PSP's velocity remains of the same order of magnitude as that of the solar wind. To improve the realism and accuracy of comparisons between synthetic and observed data, it is therefore necessary to implement a more realistic PSP trajectory.
 
For this purpose, we selected the real PSP trajectory during its first closest approach to the Sun on 24 December 2024, during Encounter 22. We employ the time and coordinate sub-packages of the Sunpy library \citep[][]{sunpy_community2020} to retrieve PSP's ephemerides between 11:21:40 and 12:01:40 UT, with a temporal resolution of $20$ s. During this period, PSP had an average velocity of 191 km s$^{-1}$, primarily in the non-radial direction. The perihelion occurred at 11:41:00 UT, with coordinates: $r_{\rm{p psp}} = 9.864~\rm{R}_{\odot}$,  $\rm{lon}_{\rm{p psp}} = 52.38 ^{\circ}$ and $\rm{lat}_{\rm{p psp}} = -3.63 ^{\circ}$. Over the full time interval, PSP's trajectory spans: $r_{\rm{psp}} \in [9.864 ~  \rm{R}_{\odot}; 9.867 ~ \rm{R}_{\odot}] $, $\rm{lon}_{\rm{psp}} \in [50.54 ^{\circ}; 54.34 ^{\circ}]$, and $\rm{lat}_{\rm{psp}} \in [-3.67 ^{\circ}; -3.58 ^{\circ}]$. 

To construct a virtual PSP trajectory within the simulation, only the latitude and longitude coordinates are transformed, while both the radial distances and the spacecraft velocities are preserved. The procedure used to define the reference virtual PSP trajectory is described below, and additional virtual trajectories are presented in Appendix~\ref{app:A}.
The transformation proceeds in two steps. First, we shift the trajectory so that a reference point close to the perihelion, the point corresponding to 11:41:40 UT with coordinates ($r_0 = 9.864~ \rm{R}_{\odot}$,  $lon_0 = 52.44 ^{\circ}$, $lat_0 = -3.63 ^{\circ}$), is projected onto the corresponding target location in the simulation domain: $r_1 = 9.864~\rm{R}_{\odot}$,  $\theta_1 = 89.58 ^{\circ}$ and $\phi_1 = 3.17 ^{\circ}$. The coordinate shift is therefore: 
\begin{equation}
        \theta_{\rm{rot1}} = \theta_{\rm{traj}} + (\theta_1 - lat_0), \quad \phi_{\rm{rot1}} = \phi_{\rm{traj}} + (\phi_1 - lon_0)
\end{equation}
Next, we applied a rotation of $\sim 37^{\circ}$ (counterclockwise direction) about the fixed target point to align the trajectory with the desired orientation within the simulation. After these two transformations, the virtual PSP path spans the subdomain of the simulation domain: $r_{\rm{sim psp}} = r_{\rm{psp}} \in [9.864 \,  \rm{R}_{\odot}; \, 9.867 \, \rm{R}_{\odot}]$, $\theta_{\rm{sim psp}} \in [88.39^{\circ};\,  90.76^{\circ}]$, $\phi_{\rm{sim psp}} \in [1.67^{\circ}; \,  4.65^{\circ}]$. 

This subdomain was specifically chosen to ensure that the virtual PSP crosses the propagating torsional Alfvénic jet wavefront (illustrated in Figure~\ref{fig:overview_sb_jet}; T1 trajectory parameters in Table~\ref{tab:tab_param}). The selected time interval, $t = [11\,000$ s; $16\,000$ s], corresponds to the period when the Alfvénic front is located near $r_0= 9.864~\rm{R}_{\odot}$, thereby maximising the physical relevance of the comparison.

Finally, unlike in \citetalias{Touresse24}, where the H$\beta$ simulation was run at a temporal resolution of $100$ s, we partly reran the simulation with a $20$ s resolution to capture finer temporal variations and to remain consistent with the cadence of the transformed PSP trajectory.

\subsection{Deflection tracking method} \label{sec:track_def}

Implementing the trajectory described in Section \ref{sec:psp_traj} (T1 trajectory parameters in Table~\ref{tab:tab_param}) provides some synthetic in-situ measurements (see right panel of Figure \ref{fig:synthetic_psp_trajectory}) studied in Section \ref{sec:mag_geo_sb}. We analyse the magnetic field lines intersected by the PSP path within the jet wavefront (e.g. diachronic magnetic connectivity, see left panel of Figure \ref{fig:synthetic_psp_trajectory}) and identify bendings in the lines embedded in the SB region compared to their initial radial direction (see Figure \ref{fig:mag_fieldlines_types}). To characterise the temporal evolution of these bendings — hereafter referred to as magnetic deflections — along a given magnetic field line, we develop a dedicated method for tracking the deflection.

In the lower radial boundary, line-tying conditions apply. Thus, we initially intended to track the evolution of these magnetic deflections using the positions of magnetic field lines from fixed footpoints at the lower radial boundary. However, intermittent, slipping reconnection occurs at low solar radii, in the vicinity of the jet generating current sheets at the source region. This alters the connectivity of individual field lines, making this approach unreliable for tracking a single line upward through the simulation domain.

To overcome this limitation, we implemented an alternative tracking strategy, based on field lines traced from positions located at the upper radial boundary. Although line-tying is not enforced at this boundary, transverse motions remain sufficiently small throughout most of the simulation (\ie before the propagating jet reaches the upper boundary) that they can be considered negligible. We therefore consider the upper-boundary starting point for the field line tracing as effectively fixed, and we explicitly verified that their displacement remains minimal. Under these conditions, field lines traced from the upper boundary can reliably be used to follow the evolution of field line connectivity. 

This upper-boundary tracking approach minimises the influence of reconnection-driven footpoint motions and enables a consistent quantification of the deflection amplitude and its propagation along the field lines inducing the SB. It thus provides a robust mean of tracing Alfvénic deflections while accounting for the practical constraints imposed by line-tying and slipping reconnection at lower radii.

With the ability to follow the same group of magnetic field lines, we can now track the evolution of the deflection along them. Because the deflection occurs tangentially to the field, we quantify it by computing, at a given snapshot time, $t$, at every point $s$ along a field line, the deviation angle, $\alpha$, between the local magnetic field direction and the radial direction:
\begin{equation} \label{eq:alpha}
    \alpha(s,t) \equiv \arctan\!\left(
    \frac{B_{\perp}(s,t)}
         {B_{\parallel}(s,t)}
\right),
\end{equation}
where $B_{\perp} = \sqrt{B_{\theta}^2 + B_{\phi}^2}$, and $B_{\parallel} = B_r$, are the perpendicular and parallel components of the magnetic field with respect to the radial direction. As the Parker spiral effect is not included in the present simulations, the background magnetic field is purely radial, and the deviation angle, $\alpha$, quantifies departures from this radial configuration rather than from a Parker-spiral geometry.

Deflections become apparent when plotting $\alpha$ along a field line, as illustrated in Figures \ref{fig:single_def_alpha_r} and \ref{fig:double_def_alpha_r}. We define a deflection as the portion of the field-line evolution in which the deviation angle, $\alpha$ progressively increases—indicating that the magnetic field line is deviating from its initial radial direction—until reaching a maximum, after which $\alpha$ decreases as the field line relaxes back toward the radial direction.

To identify the maxima of a deflection quantitatively, we use the Python function \texttt{scipy.signal.find\_peaks} with constraints on the prominence and minimum separation between local maxima identified. At a given time, the maximum value, $\alpha_{\rm max}(t)$ of the deflection along a given field line (or a subsection of it), $\mathcal{F}$, is denoted by:
\begin{equation} \label{eq:alpha_max}
\alpha_{\rm max}(t)\equiv \max_{s\ \in\ \mathcal{F}}(\alpha(s,t))\, . 
\end{equation}
At each instant $t$, we additionally compute $r_{\rm max}(t)$ for the field line of interest (or a subsection of it), that is, the radial coordinate at which the deflection angle, $\alpha$, reaches its maximum: 
\begin{equation} \label{eq:r_max}
r_{\rm max}(t)\equiv r(\alpha=\alpha_{\rm max}(t))\ . 
\end{equation}
These notations, $\alpha_{\rm max}(t)$ and $r_{\rm max}(t)$, are used throughout the rest of the article. This quantity, $r_{\rm max}(t)$, is employed to follow the temporal evolution and propagation of the maximum of deflection, which serves as a proxy for tracking the evolution of the propagating untwisting Alfvénic wave. The corresponding angle, $\alpha_{\rm max}(t)$, characterises the amplitude of the deflection and allows us to assess how it varies as the Alfvénic wavefront propagates upward.

As each deflection has a finite spatial extent, the position of its maximum may shift internally within the deflection as it evolves, independently of both the plasma flow and the global propagation of the deflection. We note that therefore the spatial evolution of the maximum deflection, $\alpha_{\rm max}(t)$,  along a field line should not be conflated with either tracking the position of a plasma element located at the point of maximum local deflection (in an Eulerian or Lagrangian sense) or with the bulk propagation of the Alfvénic wave packet itself. Before presenting the results, we stress an important distinction regarding the interpretation of the propagation velocity derived from $r_{\rm{max}}(t)$. Three distinct velocities naturally arise in this context: (i) the bulk plasma flow velocity, (ii) the propagation velocity of the Alfvénic wave packet as a whole, and (iii) the apparent velocity at which the maximum of the deflection travels within the packet. In this work, we specifically track the motion of the deflection maxima.

\section{What is the magnetic geometry of a switchback event?} \label{sec:mag_geo_sb}

\begin{figure}
  \centering
 \resizebox{\hsize}{!}{\includegraphics{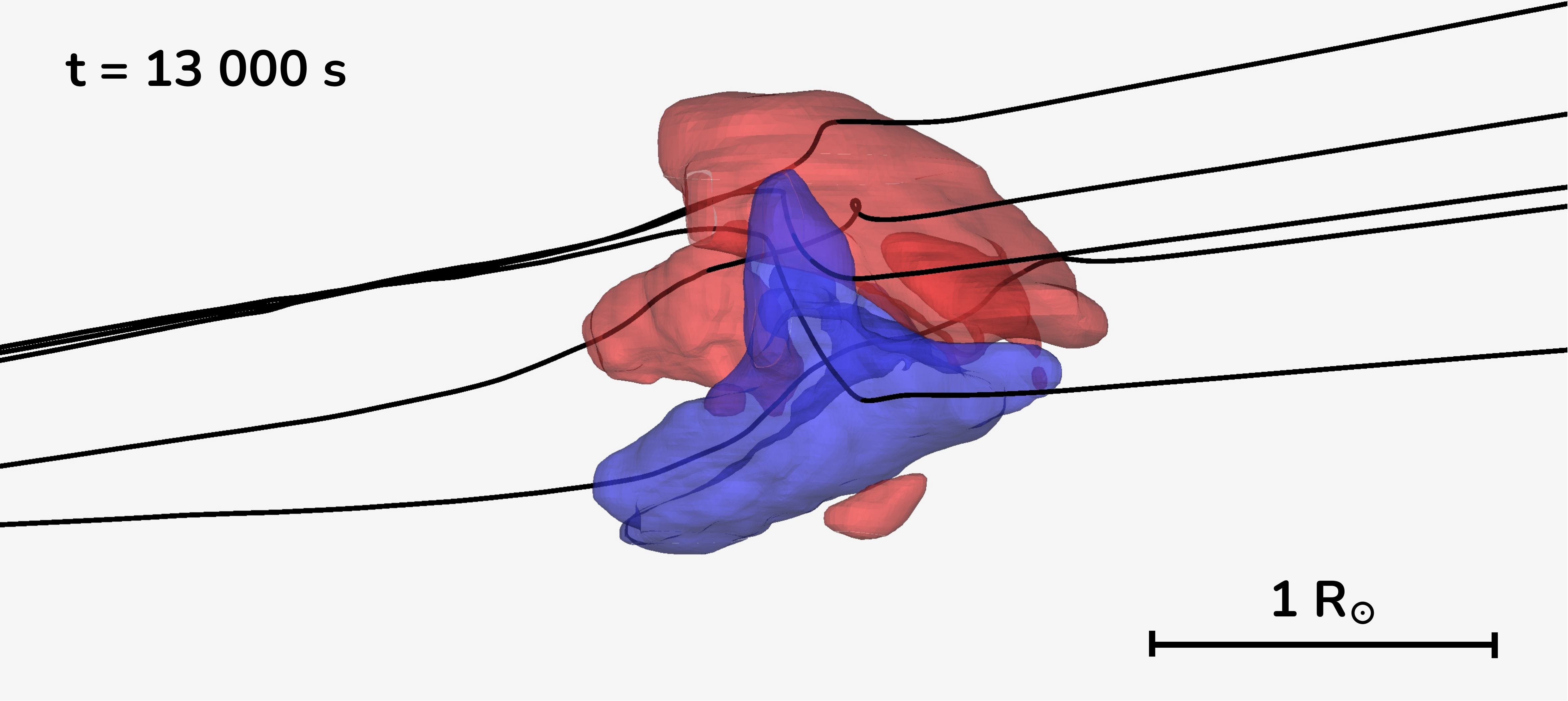}
 }
  \caption{3D visualisation of the jet-induced torsional Alfvénic structure. The blue and red isosurfaces represent $\pm 30 $ km s$^{-1}$ of the $\theta$-component of velocity ($v_{\theta}$), highlighting regions of high rotational velocities. The black lines correspond to magnetic field lines. The isosurfaces are only traced between $8.5$ R$_{\odot}$ and $10$ R$_{\odot}$.}
  \label{fig:overview_sb_jet}
\end{figure}

\begin{figure*}
  \centering
 \resizebox{0.9\hsize}{!}{\includegraphics{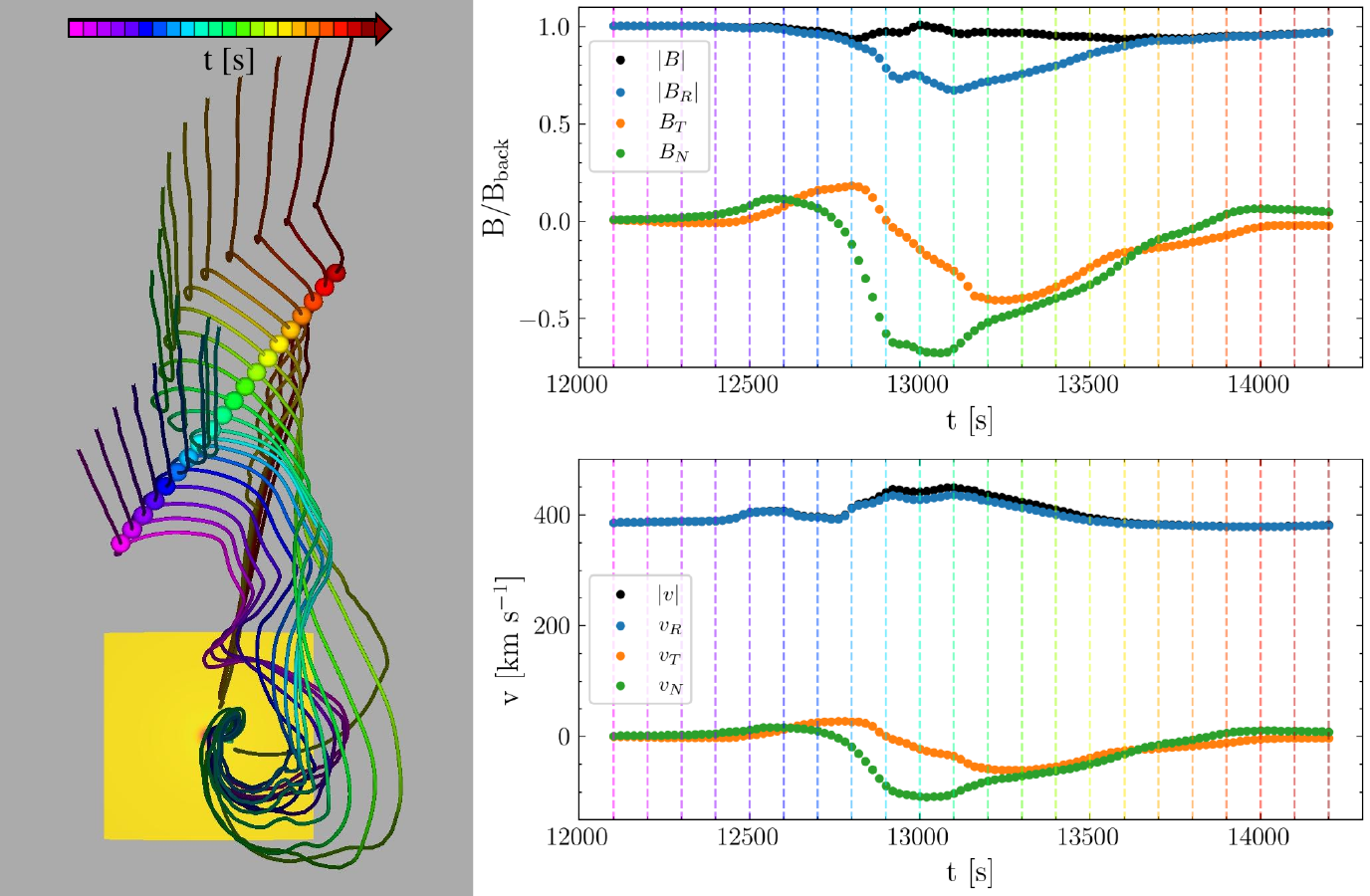}}
  \caption{Synthetic PSP trajectory, diachronic PSP magnetic connectivity and corresponding in-situ magnetic field and solar wind velocity  measurements. Left Panel : Simulated trajectory of PSP within the computational domain using positions from its closest approach (December 24, 2024). The yellow square represents the solar surface, while the coloured points indicate the virtual PSP's path, with time progression from purple to dark red. Each point corresponds to a position recorded every $dt = 100$ seconds. At each of these positions, the magnetic field line intersected by PSP is represented in the same colour as the points showing the PSP position. The figure therefore displays the diachronic magnetic connectivity of PSP.
  Right Panel: Synthetic measurements of the magnetic field and velocity along the virtual PSP's trajectory. The upper right panel presents magnetic field measurements in the RTN frame, normalised by the ambient magnetic field magnitude around the SB. The lower right panel shows the velocity measurements in the RTN frame. Both datasets were acquired every twenty seconds. The vertical dashed lines in each plot indicate the timestamps of the left panel's trajectory points, with colours corresponding to the respective data points. The \textbf{RTN} frame is defined as follows: \textbf{R} (radial) is the Sun to spacecraft unit vector, \textbf{T} (tangential) is the cross product between the Sun's spin axis and \textbf{R}, and with \textbf{N} (normal) completing the direct orthogonal frame. The associated movie is available online.}
  \label{fig:synthetic_psp_trajectory}
\end{figure*}

In our previous study \citetalias{Touresse24}, we investigated the propagation of a solar jet through the simulation domain. Consistent with prior works \citep[][]{Pariat16, Uritsky17, Roberts18}, we observed a leading Alfvénic torsional wave followed by a denser trailing jet. By analysing this jet-driven untwisting torsional magnetic wave, as if a spacecraft were traversing the simulation domain at infinite speed, we identified several signatures characteristic of SBs.

Building on these results, the present section examines the magnetic geometry of the SB region generated by the crossing of the leading Alfvénic torsional wave. The corresponding Alfvénic structure is illustrated in Figure \ref{fig:overview_sb_jet}. The selected isosurfaces reveal regions of high rotational velocity within the structure. The magnetic field lines exhibit noticeable bendings around these isosurfaces, deviating from their initial radial orientation. This provides a 3D visualisation of the Alfvénic structure of the jet.

In this study, we use an actual PSP trajectory (see Section \ref{sec:psp_traj} for details), which crosses the torsional Alfvénic wavefront in our simulation. Along this trajectory, we extract synthetic in-situ measurements of the magnetic field and plasma velocity components (see the right panel of Figure \ref{fig:synthetic_psp_trajectory}). The associated movie illustrates the temporal evolution of the synthetic trajectory and the corresponding magnetic connectivity. For clarity, the PSP synthetic position shown in the left panel of Figure \ref{fig:synthetic_psp_trajectory} is represented by a point every 100 s, colour-coded according to time. At each of these positions, we also display the diachronic magnetic connectivity, \ie the magnetic field lines sequentially intersected by the spacecraft. 

First, we focus on the right panel of Figure \ref{fig:synthetic_psp_trajectory}, which shows that, according to the standard definition of a SB, we indeed observe:
\begin{itemize}
    \item a deflection in the radial magnetic field, $B_r$, with a variation of $33$ \% (blue scattered dots, upper panel) relative to the background solar wind magnetic field,
    \item a nearly constant magnetic field magnitude, varying by only  $6$ \%  (black scattered dots, upper panel) ; the small variation at the leading boundary may correspond to a magnetic dip as reported in \citet[][]{Agapitov20, Farrell20, Larosa21}, and
    \item an increase in velocity magnitude of about $70$ km s$^{-1}$ (black scattered dots, lower panel). This corresponds to an increase of approximately 18\% above the local background solar wind velocity. 
\end{itemize}

The measured radial velocity, $\delta v_r = 70~\mathrm{km\,s^{-1}}$, is slightly larger than the Alfv\'enic expectation $(\delta B_r / B)\, v_A \approx 52~\mathrm{km\,s^{-1}}$, giving a ratio of $\delta v_r / [(\delta B_r / B)\, v_A] \approx 1.35$. Since this ratio is slightly above 1, the perturbation is weakly nonlinear, which may lead to wave steepening or the development of compressible effects.

We compute the deflection angle, $\alpha$, between the magnetic field and the radial direction, which reaches a maximum of $\alpha = 46^{\circ}$ along the trajectory of our virtual spacecraft. 

These synthetic in-situ measurements can be compared with the measurements shown in the bottom right panel of Figure~9 of \citetalias{Touresse24}. Both the maximum deflection angle and the magnetic field variation are smaller in the present study than in \citetalias{Touresse24} ($\alpha_{\rm max} = 46^{\circ}$ vs.\ 62$^{\circ}$, and 6\% vs.\ 12\%, respectively). Moreover, the SB boundaries are gradual in our synthetic data, whereas they appear sharper in \citetalias{Touresse24}; such smooth transitions are, nonetheless, often observed in in-situ measurements \citep[\eg][]{Choi25}.

We can also place the simulated event in the context of the statistical properties of switchbacks observed by Parker Solar Probe. Using the parametrisation of \citet[][]{DudokdeWit20}, a deflection angle is of $46^{\circ}$ corresponds to $z \simeq 0.15$, which corresponds to the bulk of SB events in the histogram reported in the Figure 5 of \citet[][]{Badman26}. The simulated event is therefore representative of commonly observed switchback deflection angles. Regarding duration, we estimate the simulated SB to last approximately 15 minutes (900 s). This duration lies toward the upper end of the observed duration distribution (right panel of Figure 7 of \citet[][]{Badman26} or Figure 6 of \citet[][]{DudokdeWit20}). It shall be noted that this switchback duration is a prediction of our model rather than a constraint imposed by the simulation setup, as it results from the externally imposed spacecraft’s trajectory through the simulation domain.

The objective of the present comparison is not to reproduce the observed statistical distributions of switchback properties. Rather, it shows that the simulated structure falls within the range of events observed by Parker Solar Probe. The present event may therefore be regarded as a realistic example of a relatively long, without being extreme, switchback-like structure.

The left panel of Figure \ref{fig:synthetic_psp_trajectory} provides insight into the magnetic field lines that the synthetic PSP would cross. Within the simulation time interval, $t \, \in$ [$12\,700$ s;  $13\,600$ s] - corresponding to the SB signature, we identify two distinct types of magnetic field lines. Some exhibit bending and twisting, while others remain roughly radial except in the vicinity of the probe’s trajectory, where they display localised twisting.

\begin{figure*}
  \centering
 \resizebox{\hsize}{!}{\includegraphics{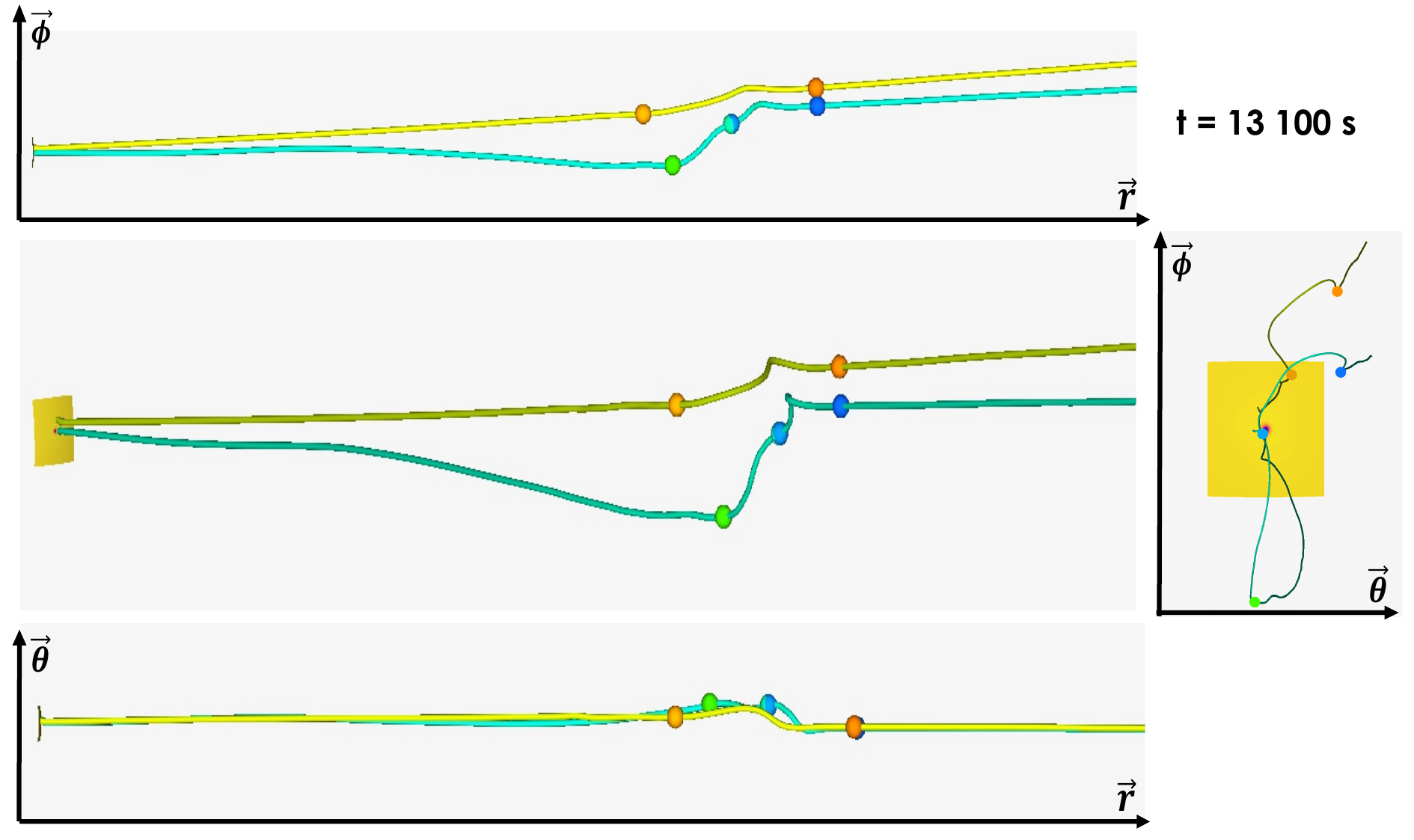}
 }
  \caption{Visualisation of two distinct magnetic field lines within the switchback shown in Figure \ref{fig:synthetic_psp_trajectory} at the same time: $t = 13\, 100$ s. 
  The solar surface is represented by the yellow square. The yellow magnetic field line corresponds to the structure crossed by PSP at $t = 13\,500$ s  in Figure \ref{fig:synthetic_psp_trajectory}, while the cyan line represents the one encountered at $ t = 13\,100$ s in Figure \ref{fig:synthetic_psp_trajectory}. The yellow field line exhibits a single deflection from the radial direction, marked by light and dark orange points indicating the start and end of it. In contrast, the blue field line undergoes two major deflections from the radial direction, with their respective start and end points shown by green, cyan, and dark blue markers. The three panels provide different perspectives of the magnetic field topology. The top panel presents the trajectory in the ($r$, $\phi$) frame. The bottom panel displays the view in the ($ r$, $\theta $) frame, while the right panel provides an alternative ($\theta$, $\phi$) perspective.}
  \label{fig:mag_fieldlines_types}
\end{figure*}

For the remainder of the article, we focus on two representative magnetic field lines shown in Figure \ref{fig:mag_fieldlines_types}. The first is the cyan-green line, corresponding to the line intersected by the virtual spacecraft at $13\,100$ s (see left panel of Figure \ref{fig:synthetic_psp_trajectory}). This magnetic field line exhibits two bendings, \ie, two deflections from the radial direction. The second is the yellow line, intersected by the virtual spacecraft at $13\,500$ s (see left panel of Figure \ref{fig:synthetic_psp_trajectory}). This magnetic field line displays one deflection. Together, these two field lines illustrate the two distinct morphologies identified within the simulated SB event.

The double-deflection field line (cyan–green) shows two distinct magnetic bends: the first, located between the green and cyan markers in Figure \ref{fig:mag_fieldlines_types}, occurs predominantly in the $\phi$ direction, while the second, delimited by the blue and dark-blue markers in Figure \ref{fig:mag_fieldlines_types}, involves deflections in both $\theta$ and $\phi$. 
Hereafter, we refer to the leading deflection (respectively trailing), when discussing the properties and dynamics of the deflection located further up (respectively further down), relative to the solar surface, along the double-deflection magnetic field line.  
By contrast, the second field line (yellow) features a single magnetic deflection, indicated by the orange and dark orange dots in Figure \ref{fig:mag_fieldlines_types}. This deflection occurs simultaneously in both $\theta$ and $\phi$.

In the following figures, orange is used for the single-deflection case. For the double-deflection case, blue and green denote the leading and trailing deflections from the radial direction, respectively.

These findings indicate that, even within a single SB event, magnetic field lines can exhibit a diverse range of geometries. Having identified a specific magnetic structure for the SB, we uncover two distinct morphologies of magnetic field lines within it: (i) field lines exhibiting a single deflection and (ii) field lines with two successive deflections.

With this classification in place, we now investigate the temporal and radial evolution of each type of field line. Do all deflections follow the same progression over time and space, or do they evolve differently? By tracking the behaviour of both categories, we aim to determine whether a universal pattern governs the evolution of these deflections. 

\section{What is the evolution of the deflection in time/radius ?} \label{sec:evol_def}

\begin{figure*}
\sidecaption
  \includegraphics[width=12cm]{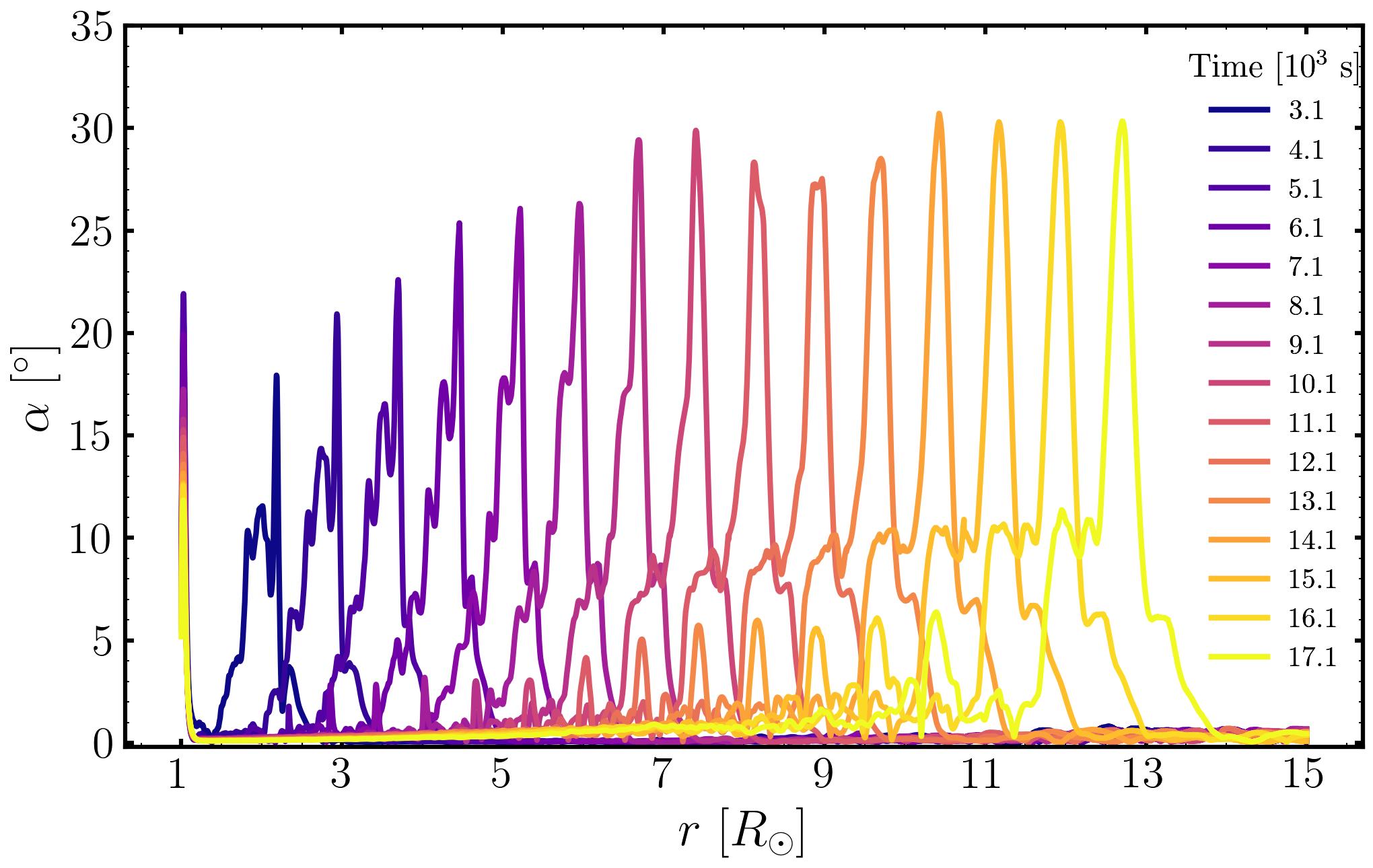}
  \caption{Spatial and temporal evolution of the angle ($\alpha$) between the magnetic field line exhibiting a single deflection (yellow line in Figure \ref{fig:mag_fieldlines_types}) and the radial direction, computed from a fixed point at the top of the simulation domain (see Section \ref{sec:track_def}). The curves show the spatial profile of $\alpha$ at times from $3\,100$~s to $17\,100$~s in steps of $1\,000$~s, with the plasma colormap indicating the temporal progression. The associated movie is available online.}
  \label{fig:single_def_alpha_r}
\end{figure*}

To analyse the evolution of the deflection along a given magnetic field line over time, a useful diagnostic is the angle formed between the magnetic field line and the radial direction  
($\alpha(s) = \arctan\!(B_{\perp}(s)/B_{\parallel}(s))$, see Equation \ref{eq:alpha}). Figure \ref{fig:single_def_alpha_r} shows the spatial evolution of this angle along the magnetic field line at various time steps for the magnetic field line case that exhibits a single deflection (yellow line of Figure \ref{fig:mag_fieldlines_types}). The corresponding temporal evolution is available in the associated movie.

Figure \ref{fig:single_def_alpha_r} exhibits a coherent structure propagating radially outward with time. For instance, at $t = 17\,100 $ s, it spans the radial range $9.9 - 13.9 \, \text{R}_{\odot}$. The structure consistently contains a prominent primary peak. At  $t = 17\,100 $ s, the peak is centred around 12.7 $\text{R}_{\odot}$ with a radial width of approximately 0.5 $\text{R}_{\odot}$. Two distinct components surround this sharp maximum: a leading packet located ahead of the peak (between 12.9 and 13.9 $\text{R}_{\odot}$) and a trailing packet extending behind it (between $9.9\, \text{R}_{\odot}$ and $12.4\, \text{R}_{\odot}$).

The leading packet appears as a modest enhancement preceding the dominant peak, whereas the trailing region exhibits a more complex and time-dependent substructure.  At $t = 17\,100 $ s, this trailing region comprises a secondary enhancement immediately behind the primary peak (between $11.6\, \text{R}_{\odot}$ and $12.4\, \text{R}_{\odot}$) and a smaller, more distant feature located between $9.9\, \text{R}_{\odot}$ and $10.6\, \text{R}_{\odot}$.

At earlier times (up to $8\, 100$ s, corresponding to approximately $6.2\, \text{R}_{\odot}$), a secondary peak is observed adjacent to the primary one. This secondary structure progressively merges with the dominant peak or with the rest of the trailing structure as the system evolves. Notably, the growth of the main peak is accompanied by an increase in the amplitude of the trailing secondary structures, while the leading packet gradually weakens over time (from $t = 8\, 100$ s onward).

We estimate the maximum angle of the deflection ($\alpha_{\rm{max}}$, Equation \ref{eq:alpha_max}) and its radial position ($r_{\rm{max}}$, Equation \ref{eq:r_max}) for each time step using the method described in Section \ref{sec:track_def}.

In Figure \ref{fig:single_def_alpha_r}, we can see that $\alpha_{\rm{max}}$ increases rapidly going from $18^\circ$ (at t = $3\,100$ s, $r_{\rm{max}}$ = $2.2\,\text{R}_{\odot}$) to $30^\circ$ (at t = $10\,100$ s, $r_{\rm{max}}$ = $7.4\,\text{R}_{\odot}$). After that, $\alpha_{\rm{max}}$ exhibits a slight oscillation around $28^\circ$. In the following section, Section \ref{sec:intensity_def}, we quantify how the magnitude of the deflection evolves with both time and radial distance.

These observations motivate further investigation into whether the two deflections along the cyan magnetic field line exhibit similar temporal and radial evolutions.

\begin{figure*}
\sidecaption
  \includegraphics[width=12cm]{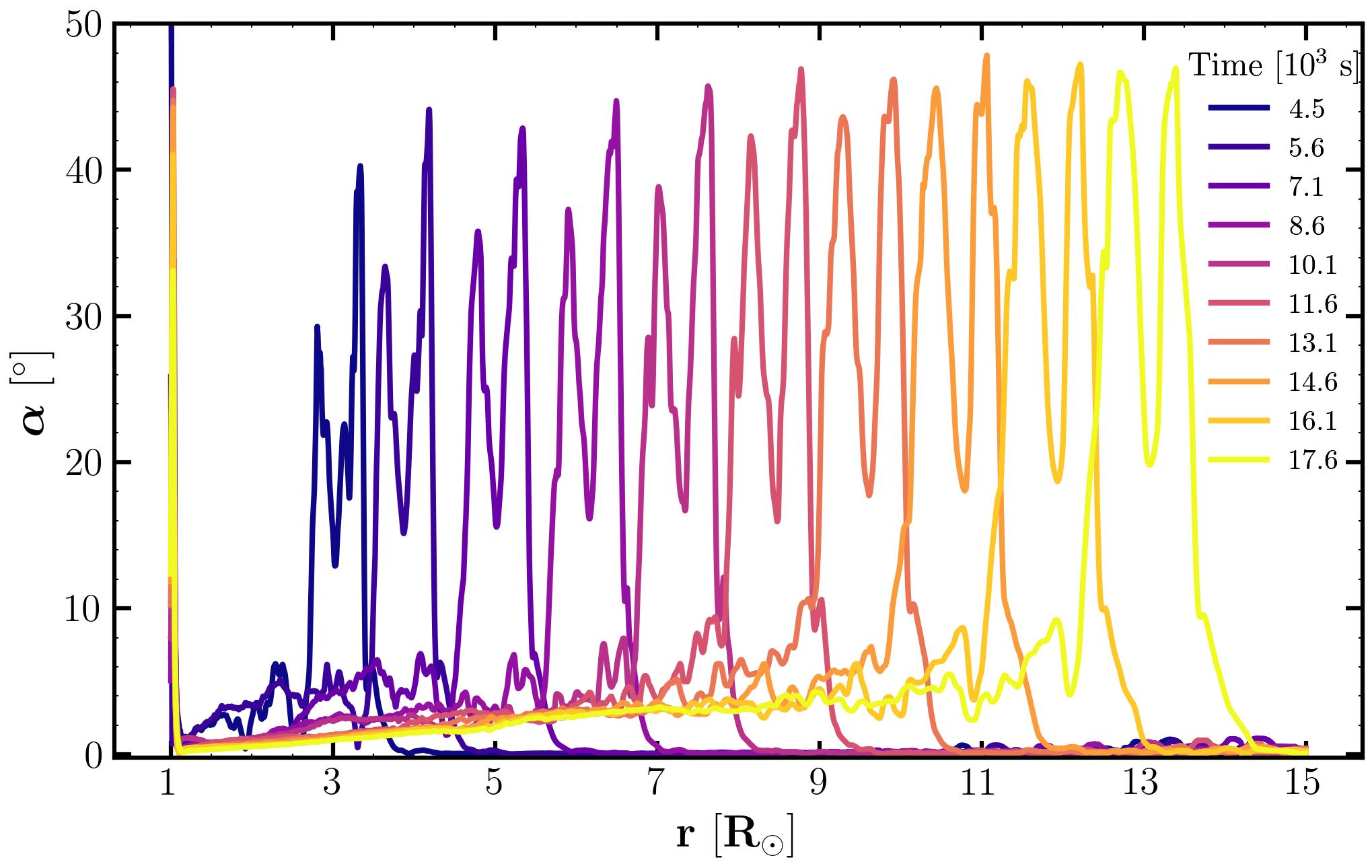}
  \caption{Spatial and temporal evolution of the angle ($\alpha$) between the magnetic field line exhibiting a double deflection (cyan line in Figure \ref{fig:mag_fieldlines_types}) and the radial direction, computed from a fixed point at the top of the simulation domain (see Section \ref{sec:track_def}). The curves show the spatial profile of $\alpha$ at times from $4\, 500$ s to $17\, 600$ s in steps of $1\,500$ s, except between $4\, 500$ s and $5\, 600$ s, with the plasma colormap  indicating the temporal progression. The associated movie is available online.}
  \label{fig:double_def_alpha_r}
\end{figure*}

Figure \ref{fig:double_def_alpha_r} shows the spatial evolution of the angle between the magnetic field line and the radial direction along a field line exhibiting two deflections  (cyan line of Figure \ref{fig:mag_fieldlines_types}) at various time steps. The corresponding temporal evolution is available in the associated movie. The plot reveals two distinct peaks propagating radially outward. At $t = 8\,600$ s, the leading peak is at $r_{\rm{l}} = 6.5 \,\text{R}_{\odot}$ and the trailing peak is located at $r_{\rm{t}} = 5.9 \,\text{R}_{\odot}$, with a smaller wave packet following these structures, between $5$ and $5.6 \,\text{R}_{\odot}$.  The leading peak is narrow and sharp, with its magnitude increasing gradually until $r \sim 9 \, \text{R}_{\odot}$ (or $t = 11\, 600$ s). Then, the magnitude of the peak oscillates around a roughly constant value ($47 ^{\circ}$).
The trailing peak grows more rapidly, reaching nearly the same magnitude as the leading peak: $47^{\circ}$  at $t = 17\,600$ s. The magnitude of the smaller packet also increases as it propagates.

The evolution of the leading peak closely resembles the single-deflection case, while the trailing peak continues to grow beyond the saturation point of the single deflection, ultimately reaching angles comparable to the leading peak. Overall, the double-deflection magnetic field line exhibits higher maximum deflection angles, reaching up to $48^{\circ}$. The detailed analysis of the evolution of the maximum deflection magnitude is presented in Section \ref{sec:intensity_def}.

We discard deflections occurring close to the solar surface. They are not propagating features but are caused by the natural curvature of the magnetic field lines due partly to the small-scale polarity. This behaviour, shown near $r \approx 1 \, \rm{R}_{\odot}$ in Figures \ref{fig:single_def_alpha_r} and \ref{fig:double_def_alpha_r}, is unrelated to the Alfvénic deflection that we aim to track.

Based on a first look at the two figures, they appear to indicate that the peaks are nearly equidistant, which suggests that they propagate at a roughly constant velocity.

These observations motivate a detailed investigation into the dynamics of the deflections. In particular, we aim to determine whether the deflections propagate at a constant velocity and, if so, to quantify this velocity.

\begin{figure}
  \centering
 \resizebox{\hsize}{!}{\includegraphics{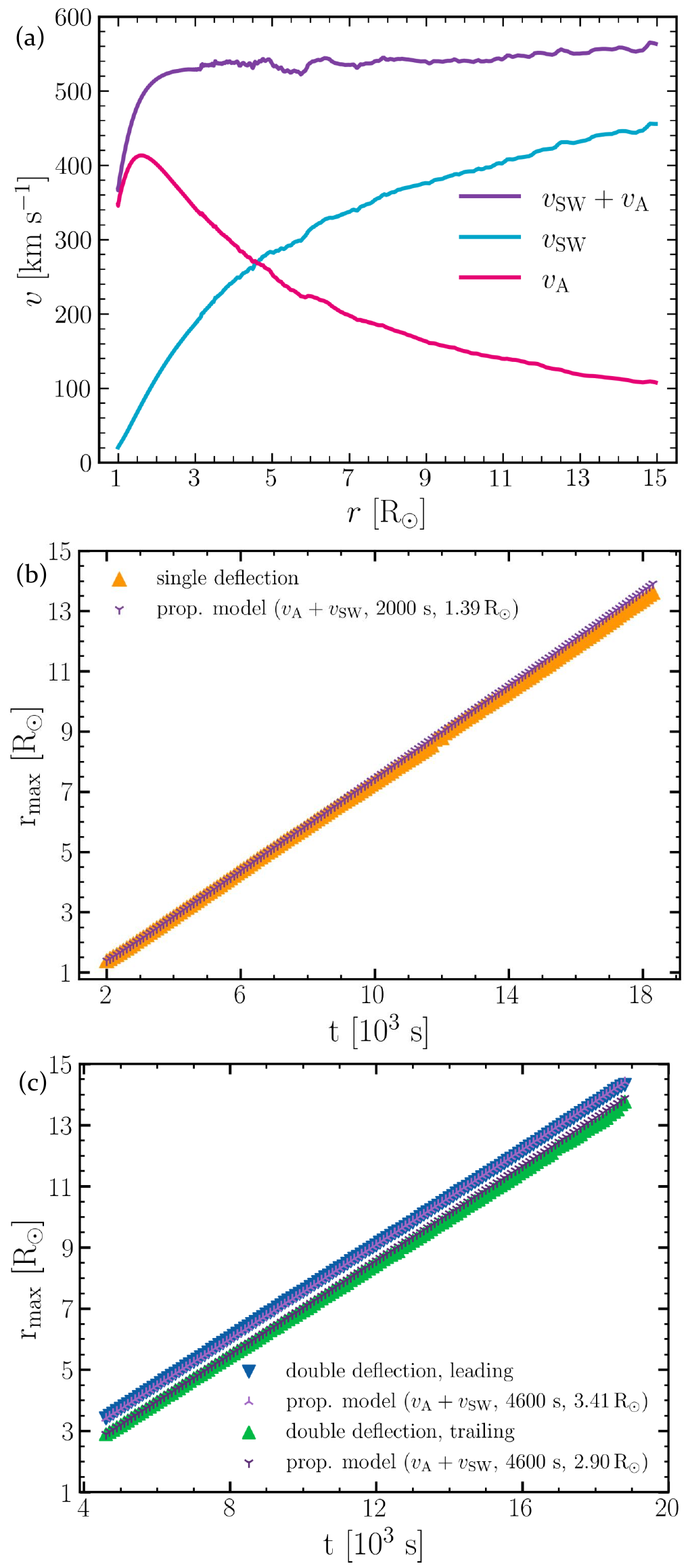}
 }
  \caption{Comparison of the measured deflection propagation velocity with the propagation velocity of an Alfvénic test structure.
  (a)  Radial profiles of the solar wind velocity, $v_\mathrm{SW}$ (blue), the Alfvén velocity, $v_\mathrm{A}$ (magenta), and their sum $v_\mathrm{sum} = v_\mathrm{SW}+v_\mathrm{A}$ (violet), obtained from the background plasma at $\theta = 97^{\circ}$, $\phi = 7^{\circ}$.
  (b) Temporal evolution of the radial position of the maximum deflection in the single deflection case (orange triangles). The purple tri-downs show the predicted trajectory of a test structure launched from the same initial position and propagated at $v_\mathrm{SW}+v_\mathrm{A}$.
  (c) Temporal evolution of the radial positions of the leading (blue triangles) and trailing (green triangles) deflection maxima in the double-deflection case. The corresponding predicted trajectories at $v_\mathrm{SW}+v_\mathrm{A}$, initialised at each maximum, are shown by the light purple (leading) and dark purple (trailing) tri-downs or tri-ups.}
  \label{fig:va_vsw}
\end{figure}

To this end, we first identify the maxima of a deflection, $\alpha_{\rm{max}}$, using the quantitative method described in Section \ref{sec:track_def}. At each time step, we record the corresponding radial position, $r_{\rm{max}}$ (see Equation \ref{eq:r_max}). For the single deflection case, this yields one $r_{\rm{max}}$ position per time step. In the double-deflection case, two values are obtained, corresponding to the radial positions of the leading and trailing maxima, respectively. To extract those radial positions, we apply the method described in Section \ref{sec:track_def} separately to the two portions of the magnetic field line associated with each peak, using the spatial extents identified previously.

For the single deflection, the temporal evolution of $r_{\rm{max}}$ is nearly linear (Figure \ref{fig:va_vsw} (b)). A linear regression yields a propagation velocity of the maximum of approximately $525.0 \, \mathrm{km\,  s}^{-1}$, with a Pearson correlation coefficient $r = 0.99997$  and a standard error of  $\sigma_m = 0.3\,\mathrm{km\,s^{-1}}$, indicating a high degree of linearity. 

In the double-deflection case, the temporal evolution of the radial positions of both maxima (leading and trailing) also evolves nearly linearly (Figure \ref{fig:va_vsw} (c)). The linear regression of the maximum at the leading deflection yields a propagation velocity of $536.3 \,\mathrm{km\,s^{-1}}$, with a Pearson correlation coefficient of $r = 0.99998$ and a standard error of $\sigma_m = 0.2$ km s$^{-1}$. The maximum of the trailing deflection propagates slightly slower, at $529.3$ km s$^{-1}$ ($r = 0.99997$ and $\sigma_m = 0.3$ km s$^{-1}$). 
The nearly perfect linearity of all curves demonstrates that the maxima of the deflections propagate coherently, at essentially uniform velocities. 

To assess the physical meaning of the measured propagation velocity of the maxima, we compared it with the characteristic background velocities. Figure \ref{fig:va_vsw} (a) shows radial profiles of the solar wind velocity (blue), the Alfvén velocity (pink), and their sum (purple), computed from a radial probe at $\theta = 97 ^\circ$, $\phi = 7 ^\circ$ at $t = 100$ s. These profiles are identical to those presented in the bottom panel of Figure 4 in \citetalias{Touresse24}. This probe provides background values away from the deflection positions and the jet. 

The sum of the Alfvén and solar wind velocities (hereafter referred to simply as the sum velocity) rises steeply near the solar surface and becomes nearly constant beyond r $\approx 1.7$ $\rm{R}_{\odot}$. Between r = 1.7 $\rm{R}_{\odot}$ and r = 15 $\rm{R}_{\odot}$, its mean value is $v_\mathrm{SW}+v_\mathrm{A} = 536 \, \pm 24 \, \mathrm{km} \, \mathrm{s}^{-1}$, which agrees well with the measured slope of the deflection maxima in Figure \ref{fig:va_vsw} (b).
A linear regression over this radial range (r = 1.7 $\rm{R}_{\odot}$ - 15 $\rm{R}_{\odot}$), yields a slope of $2.5  \, \mathrm{km} \, \mathrm{s}^{-1}$ and a Pearson correlation coefficient of $r = 0.88$. This corroborates that the sum velocity increases weakly with distance. 
Overall, this sum velocity is consistent with the apparent propagation velocity inferred from the maxima of the deflections. 

To further assess whether the measured propagation velocity of the deflection maxima is compatible with an Alfvénic behaviour, we constructed a simple propagation (toy) model.

Starting from the kinematic relation: 
\begin{equation}
    \frac{dr}{dt} = v_{\rm{model}},
\end{equation}

we integrate from the initial time  $t_{i}$ to the current time $\mathcal{T}$ and from the initial radius $r_{i}$ to the current radius $\mathcal{R}$:

\begin{equation}
     \int_{r_{i}}^{\mathcal{R}} dr = \int_{t_{i}}^{\mathcal{T}} v_{\rm{model}} \,  dt
\end{equation}

Because we have discrete radial positions (and also discrete times), the propagation model reduces to the discrete sum: 

\begin{equation}
     \mathcal{R}(v_{\rm{model}}, t_{i}, r_{i}) = \sum_{t_i : \Delta t : \mathcal{T}} v_{\rm{model}} \,  \Delta t + r_{i}
\end{equation}

We track the evolution of the deflection along the two magnetic field lines from the chosen initial time: $t_{i, s} = 2\, 000$ s for the single-deflection case and from the initial time: $t_{i, d} = 4\, 600$ s for the double-deflection case. 
We thus initialise the model using the measured radial positions of the deflection maxima. For the single-deflection case: $r_{\rm{max}} = 1.39 \,  \rm{R}_{\odot}$ at t = $2\, 000$ s. For the double-deflection case: $r_{\rm{max l}} = 3.41 \,  \rm{R}_{\odot}$ (leading deflection), $r_{\rm{max t}} = 2.90 \, \rm{R}_{\odot}$ (trailing deflection) at t = $4\, 600$ s. 

Using the background radial profile of the sum velocity: $v_{\rm{model}} = v_\mathrm{SW}+v_\mathrm{A}$ from Figure \ref{fig:va_vsw} (a), we compute the expected trajectories of test structures propagating outward at an Alfvénic velocity. These predicted radial positions  are shown in Figure \ref{fig:va_vsw} (b, c) with the purple markers: triangles down for the single-deflection test structure, light purple triangles up and dark purple triangles down for the leading and trailing test structures in the double-deflection case. 
 
For the single deflection case (Figure \ref{fig:va_vsw} (b)), the modelled trajectory closely matches the observed evolution of the maximum of the deflection at early times. At later times, the model propagates slightly faster than the measured maximum of the deflection: At $t = 18\,300$ s, there is a spatial difference of $0.36 \,  \rm{R}_{\odot}$. This suggests either a mild deceleration of the maxima, a departure of the local plasma properties from the background profile, or simply the limitations of our very idealised propagation model.

In the double-deflection case (Figure \ref{fig:va_vsw} (c)),  the maximum of the leading deflection follows the modelled trajectory remarkably well across the entire time range (at $t = 18\, 800$ s, there is a spatial difference of $0.17  \,  \rm{R}_{\odot}$). For the maximum of the trailing deflection, the two sets of points closely match at early times, but a small divergence appears at later times (at $t = 18\, 800$ s, there is a spatial difference of $0.19  \,  \rm{R}_{\odot}$), with the model structure propagating faster. In addition to the previous explanations for the single-deflection, another plausible explanation is that the leading deflection perturbs the plasma environment, modifying the local background conditions experienced by the trailing one.

As discussed in Section \ref{sec:track_def}, the velocities reported from the linear regressions should therefore be interpreted as the apparent motion of the deflection maxima, rather than as the bulk propagation velocity of the entire Alfvénic perturbation. The strikingly linear evolution of $r_{\rm max}(t)$ shown  in Figure \ref{fig:va_vsw} (b) and (c) indicates that the position of the maximum propagates outward at an approximately constant rate over the interval considered. Because of the location of the maximum is straightforward to identify and track throughout the simulation, its motion provides a useful and robust proxy for the propagation of the deflection.
This distinction naturally accounts for the small differences observed between the theoretical Alfvénic propagation velocity $v_{sw} + v_A$ and the measured motion of the maxima in the present simulation. Despite these differences, the close agreement between the apparent (measured) velocity of the maxima and the independently computed $v_{\rm SW} + v_{\rm A}$ profile, evaluated from a background probe located away from the deflection, supports an Alfvénic interpretation of the propagating structures.

In summary, both the single- and double-deflection cases exhibit maxima that propagate radially outward at nearly constant velocity, in close agreement with the sum of the solar wind and Alfvén velocities. This confirms that the deflections behave as propagating Alfvénic structures advected by the solar wind. While the radial positions of the maxima evolve uniformly, their magnitudes follow different trends: the single and leading deflections saturate beyond $\sim 7 \, \rm{R}_{\odot}$, whereas the trailing one continues to grow up to $\sim 12 \, \rm{R}_{\odot}$, where the maximum deflection angle begins to plateau at $ \sim 46^\circ$ . These contrasting evolutions of the magnitude of the deflections demonstrate that it does not simply follow their propagation and therefore warrants further investigation. This motivates the analysis presented in the next section.

\section{How does the magnitude of the deflection evolve in time/radius ?} \label{sec:intensity_def}

\begin{figure*}
  \centering
 \resizebox{\hsize}{!}{\includegraphics{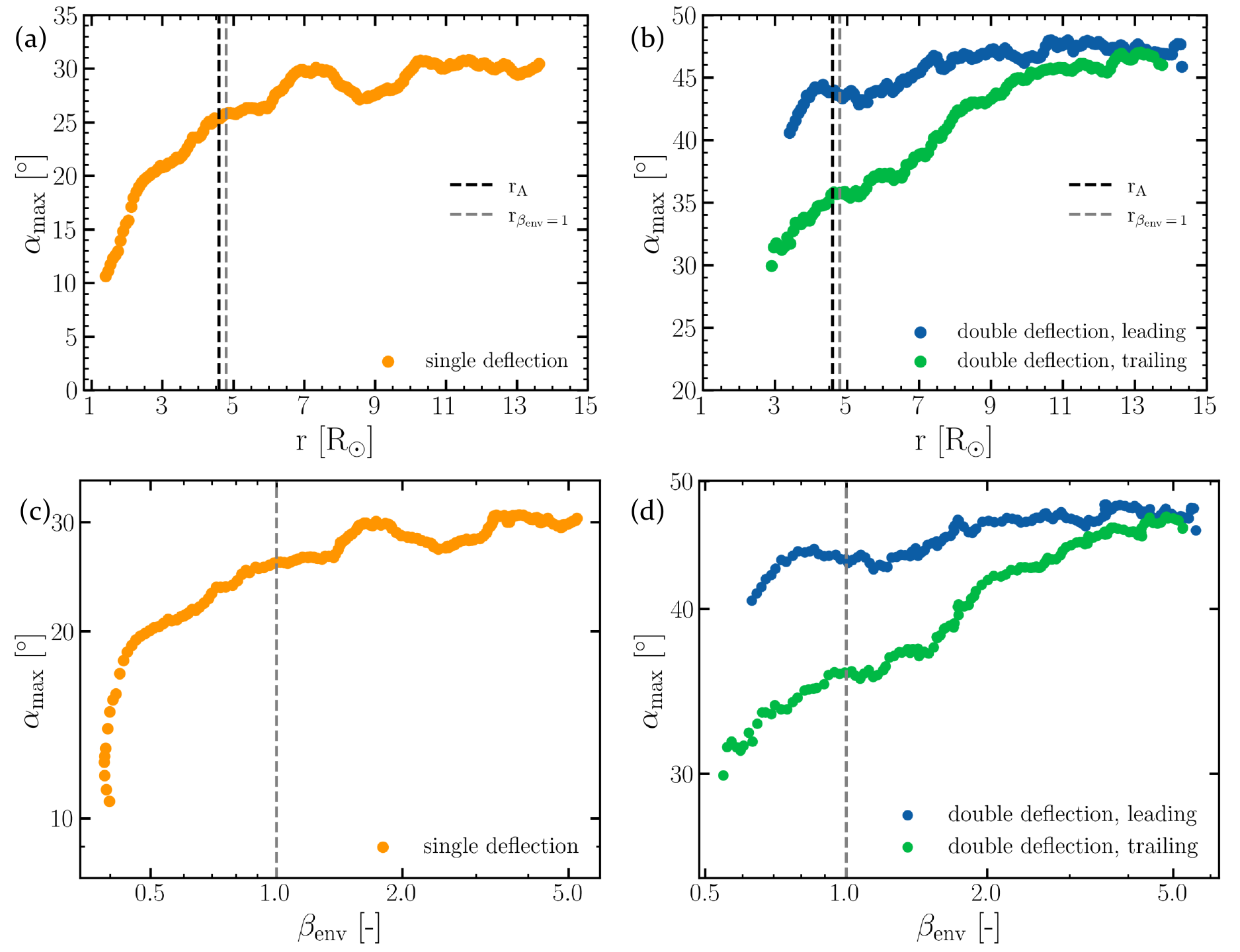}
 }
  \caption{Evolution of the maximum deflection angle along magnetic field lines with one deflection for the single-deflection case (yellow field line) and with two deflections for the double-deflection case (cyan field line).
Panels (a) and (b) show the radial evolution of the maximum angle of the deflection, while panels (c) and (d) show their evolution as a function of the plasma $\beta$ of the surrounding environment.
For the single-deflection case, the maximum angle of deflection is shown in orange; for the double-deflection case, the leading and trailing maximum angle of deflection are shown in blue and green, respectively.
In panels (a) and (b), the Alfvén radius (black dotted line) and the radius where the environmental plasma $\beta = 1$ (grey dotted line) are indicated.
In panels (c) and (d), only the $\beta_{\mathrm{env}} = 1$ location is shown (grey dotted line).}
  \label{fig:max_alpha_r_beta}
\end{figure*}

To further characterise and understand the physical nature of these structures, we investigate how the magnitude of the deflections evolves as they propagate outward. We therefore examine the spatial and temporal evolution of $\alpha_{\rm{max}}$.  Beyond simply complementing the propagation velocity analysis, we seek to determine whether the deflection magnitude can continue to grow indefinitely—potentially reaching a full magnetic reversal—or whether it instead saturates at some radial distance. If such a saturation occurs, it becomes essential to identify the mechanisms responsible for limiting the deflection magnitude in order to understand the dynamics and stability of these deflections, whose maxima have been shown to propagate like an Alfvénic wave embedded in the solar wind (see Section \ref{sec:evol_def}).

Given that the maxima of the deflections, $\alpha_{\rm{max}}$, propagate at nearly uniform velocities (Figure \ref{fig:va_vsw} (b) (c)), there is a linear relationship between their radial position and time. Consequently, the evolution of the value of the deflection maximum is almost invariant under the choice of radius or time as the reference quantity. For clarity and consistency, hereafter we choose to analyse the evolution of the value of the maximum angle of the deflection, $\alpha_{\rm{max}}$, as a function of the radial position (Figure \ref{fig:max_alpha_r_beta} (a) (b)). \\

We first focus on the single-deflection magnetic field line (the yellow magnetic field line in Figure \ref{fig:mag_fieldlines_types}). The radial evolution of the maximum deflection magnitude is shown in panel (a) of the Figure \ref{fig:max_alpha_r_beta}. The maximum angle of deflection increases from $10.6^{\circ}$ at the initial radial distance of $1.39 \, \text{R}_{\odot} $ to $30.8^{\circ}$ at $13.63 \, \text{R}_{\odot}$, illustrating a significant growth in the maximum deflection magnitude over the observed radial range.

A closer inspection of panel (a) of Figure \ref{fig:max_alpha_r_beta} reveals that the magnitude of the maximum of the deflection along the field line increases rapidly up until approximately  $3\, \text{R}_{\odot}$. Beyond this distance, the growth becomes more gradual, marking a transition from the initial phase of rapid increase. From about $7\, \text{R}_{\odot}$ onward, the maximum deflection magnitude appears to oscillate around a mean value of $29.5^{\circ}$, suggesting that the structure reaches a quasi-stationary state in terms of its maximum magnitude with only minor fluctuations around this value.

For reference, two characteristic radii of the stratified background atmosphere and solar wind are indicated in the panels (a) and (b) of Figure \ref{fig:max_alpha_r_beta}: the Alfvén radius (black vertical dashed line) and the radius at which the background plasma $\beta$ reaches unity, $\beta_{env}  = 1$ (gray vertical dashed line). These quantities correspond to those of the ambient plasma environment surrounding the SB event, rather than the local quantities within the SB event, where the deflection is. Notably, the evolution of the maximum deflection magnitude exhibits distinct changes in behaviour near these two characteristic radii. This suggests that the evolution of the deflection magnitude is closely linked to the radial variation of the background plasma and magnetic field properties.

We now turn to the double-deflection field line type (cyan field line in Figure \ref{fig:mag_fieldlines_types}) by examining panel (b) of Figure \ref{fig:max_alpha_r_beta}. Both deflection maxima exhibit an overall increase in magnitude with heliocentric distance. The maximum value of the angle of the leading deflection (plotted in blue) grows from $40.6^{\circ}$ at $3.41 \, \text{R}_{\odot}$ to a maximum of $47.9^{\circ}$, while the magnitude of the maximum of the trailing deflection increases from $29.9^{\circ}$ at $2.92 \, \text{R}_{\odot}$ to $47.9^{\circ}$.

For the leading deflection, the maximum deflection angle increases rapidly below both the  Alfvén radius and the radius at which the background plasma $\beta$ reaches unity, $\beta_{env}  = 1$. This phase is followed by a more moderate growth up to about $7\, \text{R}_{\odot}$. After that, the evolution becomes nearly flat, indicative of a saturation phase. Overall, this evolution closely resembles the one observed for the single-deflection case. In contrast, the trailing deflection (plotted in green) displays a simpler evolution: a rapid growth phase followed directly by a phase of weak growth beyond approximately $8.5\, \text{R}_{\odot}$. Notably, the crossing of plasma $\beta = 1$ does not appear to significantly affect its evolution.

The similarities between the single-deflection case and the leading deflection in the double-deflection configuration suggest that the evolution of the magnitude of the maximum of the deflection is closely linked with the radial variation of the background plasma beta. To investigate this relationship further, we analyse the evolution of the magnitude of the maximum deflection angle as a function of the background plasma $\beta$, as shown in panels (c) and (d) of Figure \ref{fig:max_alpha_r_beta}.

Focusing first on the single-deflection case (panel (c) of Figure \ref{fig:max_alpha_r_beta}), we find that at low background plasma beta values the maximum deflection angle, $\alpha_{\rm max}$, increases rapidly, from $10.6^{\circ}$ at $\beta_{\rm env}=0.40$ to approximately $21.0^{\circ}$ at $\beta_{\rm env}=0.47$. This initial growth phase is followed by a second phase in which the increase of $\alpha_{\rm max}$ becomes more gradual. Around  $\beta_{\rm env} \simeq 1.5$, the maximum deflection angle fluctuates around a mean value of $\sim 29.5^{\circ}$. We note, however, that this latter phase corresponds to plasma beta values that are not typically observed in the solar wind and is therefore unlikely to be physically realistic. Measurements from Helios and Ulysses show that the solar wind plasma $\beta$ is generally low at $\sim$ 0.3 au ($65 \, \text{R}_{\odot}$) and increases to values close to unity near 1 au \citep[][]{Matteini07}. However, PSP observations indicate that plasma $\beta$ can exceed unity, reaching values between 1 and 10 at heliocentric distances as small as $\sim 16 \, \text{R}_{\odot}$ \citep[][]{Kasper21}. 

We now consider the double-deflection case shown in panel (d) of Figure \ref{fig:max_alpha_r_beta}. For the leading deflection (blue symbols), the maximum deflection angle, $\alpha_{\rm max}$, increases rapidly up to $\beta_{\rm env}\simeq0.75$, where it reaches approximately $43^{\circ}$. Beyond this point, the growth rate decreases significantly, leading to a progressive saturation of the deflection amplitude. In contrast, the maximum angle of the trailing deflection (green symbols) does not exhibit a comparable trend. Its evolution is characterised by a more moderate increase, with indications of saturation only at very high plasma beta values, which again lie outside the range expected under realistic coronal conditions.

Overall, these results indicate that the evolution of the magnitude of the maximum angle of the deflection is not uniform but is governed by the spatial variation of the background plasma properties, in particular the plasma beta and the associated Alfvén velocity. The transition occurring near $\beta_{\rm env}\approx1$ marks a shift from magnetically dominated to thermally dominated dynamics, which appear to control both the amplification and the eventual saturation of the magnitude of the maximum of the angle of the deflection. In the single deflection case and the leading deflection in the double-deflection case, the dominant evolution of $\alpha_{\rm max}$ occurs typically in two distinct phases: one of sharp growth for $\beta_{\rm env}<<1$, and then a more moderate growth phase for $\beta_{\rm env}\sim 1$  whereas the trailing deflection appears comparatively insensitive to this transition.

Because the quantities are plotted on logarithmic axes, the evolution of the maximum deflection angle can be interpreted in terms of power-law behaviours. For the single deflection and for the leading deflection in the double-deflection case, two distinct power-law scalings are apparent, corresponding to the two growth phases identified previously. In contrast, the trailing deflection is well described by a single power-law over the explored range, indicating an overall moderate growth without a clear transition between phases.  However, we stress that this analysis is based on a limited number of magnetic field lines.

\section{Conclusion}\label{sec:conclusion}

\subsection{Summary}

The present study investigates the crossing between the Alfvénic magnetic wavefront associated with a propagating solar jet and a virtual PSP trajectory, constructed based on an actual PSP orbit, within an adaptive 3D MHD simulation. In particular, we focus on the 3D magnetic geometry of the SB region and the dynamics of the associated magnetic deflections within these structures. Our simulation reveals, for the first time, the fully 3D magnetic configuration of a consistent SB region formed within a jet-driven coronal environment. We note that the presented configuration constitutes one possible magnetic arrangement, illustrating SB formation in a solar-origin scenario. While previous observational studies have examined the global shapes or aspect ratios of SBs \citep[][]{Horbury20, Mozer20, Laker21}, their internal magnetic geometry has remained largely unconstrained. Here, we provide a direct visualisation of the magnetic field structure together with synthetic in-situ measurements along a trajectory comparable to that of Parker Solar Probe.

The simulated SB contains (at least) two distinct families of magnetic field lines: one family where field lines exhibit a single deflection and another family where two sequential deflections propagate along the field lines (see Section \ref{sec:mag_geo_sb}).
In both cases, the maximum of these deflections propagates at approximately the value: $v_{\text{SW}}+v_{\text{A}}$ of the environment, consistent with outward-travelling Alfvénic deflections embedded in the solar wind flow. This behaviour, well established in earlier theoretical studies \citep[][]{deHoffmann50, Belcher71, Barnes74, Hollweg74, Marsch87}, shows that our model naturally reproduces the quasi-Alfvénic nature of SBs reported by in-situ observations \citep[][]{Bale19, Kasper19, Horbury20}, and is therefore compatible with a solar-origin scenario for SB formation.  

As discussed in Section \ref{sec:intensity_def}, the magnitude of the maximum of the deflections is strongly modulated by the background plasma $\beta$ in the present simulation. In low-$\beta$ regions, there is a strong amplification of the magnitude of the deflection with radial distance, whereas in regions with $\beta \sim 1$, the growth of the magnitude of the deflection is more moderate.  Interestingly, in the double-deflection configuration, the trailing deflection does not follow this trend and shows a single power-law on the explored range. 

\subsection{Discussion and perspectives}

The 3D MHD simulation presented in this work offers a physically grounded framework to explore the magnetic geometry and dynamics of a SB region induced by the Alfvénic torsional wavefront of a solar jet event. Building on the quantitative results described above, we now examine the broader implications, limitations, and connections of these findings to previous observational and theoretical studies.

The main limitations of our model are similar to those discussed in Section 5.2 of \citetalias{Touresse24}, namely the simplified treatment of the coronal background, the size of the jet, the isothermal assumption, and the limitations of the near-ideal MHD paradigm. These modelling choices defined a controlled but idealised framework that may influence the long-term evolution of the simulated jet and the propagation of the associated Alfvénic deflections. Nevertheless, the model captures some essential physical processes, including magnetic reconnection and the generation of large-scale magnetic field-line deflections, making it suitable for examining the early development of SB-like structures in a solar-origin scenario. 

In this context, the present work should be viewed as a physics-based numerical investigation rather than a statistical model of the switchback population. The objective is to identify the physical processes and three-dimensional magnetic geometries capable of generating switchback-like signatures and to investigate how the resulting in situ properties depend on the spacecraft trajectory through the structure. A quantitative comparison with observed switchback distributions would require an ensemble of simulations spanning a range of jet properties, coronal environments, and impact parameters, which is beyond the scope of the present study.

The 3D geometry of the simulated SB is strongly influenced by both the orientation and the distance of the synthetic PSP trajectory relative to the jet axis or the Alfvénic wavefront (see left panels of Figure \ref{fig:synthetic_psp_trajectory} and \Figss{reverse_trajectory}{trajectory_7}).
A trajectory passing close to the jet axis samples the core of the structure and results in a broader SB region, whereas an oblique or off-axis trajectory—analogous to a secant intersecting a circle—intersects only the peripheral regions, naturally producing narrower SB signatures. As the distance of the trajectory from the jet axis increases, the width of the SB region correspondingly decreases, leading to SBs of different durations in the in-situ measurements. Moreover, the region where the radial magnetic field deviates from the radial direction forms a corkscrew-like structure that is narrow near the apex of the Alfvénic wavefront and progressively expands along the jet axis. Sharp boundaries observed in the in-situ data are associated with regions of stronger radial magnetic-field gradients along this corkscrew. Depending on whether the trajectory crosses the corkscrew while moving away from or toward the jet axis, the resulting SBs exhibit longer or shorter durations, respectively. This geometrical configuration naturally explains the observed variability in both the size (or duration) and the boundary sharpness of SBs.

The torsional Alfvénic wavefront of a single jet-like event, when sampled along different synthetic PSP trajectories, can produce a wide variety of SB signatures (see right panels of Figure \ref{fig:synthetic_psp_trajectory} and \Figss{reverse_trajectory}{trajectory_7}). These trajectories correspond to different impact parameters, implying that observed SB statistics \citep[\eg][]{DudokdeWit20, Larosa21} are shaped not only by the intrinsic magnetic structure but also by the spacecraft path through the SB region. A similar observation bias has been shown in CME studies \citep[][]{Janvier13}. Recovering the intrinsic statistical properties of SBs therefore requires disentangling these trajectory effects. In this context, the present model offers a way to deconvolve observed SB statistics from the spacecraft impact parameter. More broadly, this model provides a framework to investigate intrinsic geometrical properties of SBs, including their aspect ratio and boundary sharpness, which have also been inferred from in-situ observations \citep[][]{Horbury20, Laker21, Stumpo26}.  

In our simulation, the jet exhibits a lateral extent of approximately $6 \times 10^5$ km ($\sim 600$ Mm) at the heliocentric distance corresponding to the PSP crossing, comparable to the spatial scales inferred from time profiles of large individual SBs and consistent with values reported in previous observational studies of coronal jets \citep[][]{Sterling20} and observational studies on in-situ SBs \citep[][]{Larosa21, Laker21, Badman26}.
Above the fan-spine, the jet width is approximately 56 Mm as measured from the velocity distribution (28 Mm from the density distribution), and the characteristic scales of the underlying magnetic structure; a parasitic polarity region of $\sim$ 25 Mm  and a separatrix fan extending over $\sim$50 Mm; together place the simulated jet toward the upper end of the observed coronal jet size distribution \citep[][]{Shimojo96, Savcheva07}. In \citet[][]{Shimojo96} specifically, jet widths ranging from 3 to 100 Mm with a mean of $\sim$ 18 Mm have been reported. Because the jet propagates outward while the spacecraft traverses it, the resulting duration of the simulated magnetic switchback is shorter than would be expected for a stationary structure of equivalent spatial extent.  Jets of smaller extent would naturally produce switchback signatures with correspondingly shorter durations. This likely contributes to the simulated event falling toward the upper end of the observed switchback-duration distribution, since the jet considered here is relatively large and therefore generates a spatially more extended Alfvénic deflection region. A systematic investigation of how switchback duration scales with jet size would require an ensemble of simulations spanning a range of source-region properties and spacecraft impact parameters, and represents a natural direction for future work.

The internal magnetic geometry of a single SB appears to be more complex than that of a simple torsional structure with field lines organising into at least two configurations: single-deflection and double-deflection. The coexistence of two field-line populations suggests a more complex magnetic geometry that cannot be explained by a single torsional Alfvén wave launched by one reconnection event. Instead, it is indicative of a more intricate formation process possibly reflecting multiple or sequential reconnections at the jet base. The single-deflection field lines likely correspond to a simple torsional Alfvénic pulse generated by the untwisting of newly opened field lines following reconnection. In contrast, the double-deflection lines may represent compound magnetic structures: interacting Alfvénic fronts, the combination of kink and torsional components, or sequential reconnection episodes of differing impulsiveness. This diversity suggests that SBs are not unique topological entities but rather a continuum of Alfvénic deflections shaped by the local reconnection history, plasma environment, and propagation effects \citep[][]{Tenerani20}. 
 
Our numerical model predicts that a given observed SB may not only be the result of simple S-shaped folds of the magnetic field, as it is almost always sketched in the SB literature, but can actually consist of diverse and complex field line morphologies (see Figure \ref{fig:synthetic_psp_trajectory} and Appendix \ref{app:A}). This may explain part of the complex magnetic field variations which are very commonly seen within observed SBs. Such complexity may explain why PSP often detects SBs with varying degrees of field rotation or asymmetry \citep[][]{DudokdeWit20, Larosa21}.
While SBs boundaries have been the topic of several studies \citep[\eg][]{Farrell20, Froment21, Suen23, Bizien23}, such magnetic dynamics within the core of the SBs have so far been poorly explored. Our numerical SB model may thus become a tool of interest to further investigate the diversity of observed SB internal structure and its link with the 3D magnetic morphology of the field lines.

The propagation of the maxima of the deflection angles along the magnetic field lines at velocities consistent with $v_{\text{SW}}+v_{\text{A}}$ of the environment and with no significant deceleration or dispersion suggests that the simulated structures remain broadly linear Alfvénic responses consistent with an impulsive reconnection of the jet and the subsequent launch of untwisting torsional Alfvénic waves. The ability of these coherent deflections to persist over several solar radii implies that SB-like structures may preserve magnetic connectivity to their coronal source region, at least over the near-Sun domain, lending support to interpretations in which a substantial fraction of observed SBs correspond to outward-propagating torsional Alfvén waves launched by reconnection in the low corona.

A caveat in the interpretation of the fitted velocities is that they are derived from the trajectory of the deflection maximum rather than from a direct measurement of the propagation speed of the underlying Alfvénic perturbation. Even for a coherent, non-dispersive Alfvénic wave packet propagating at approximately $v_{\rm SW} + v_{\rm A}$, the location of the deflection maximum cannot remain tied to a fixed position within the packet. During the evolution of the structure, variations in the background properties and weak nonlinear effects may gradually modify its shape, causing the maximum to shift internally. Consequently, the velocity inferred from $r_{\rm max}(t)$ characterises the apparent motion of the intensity maximum and not necessarily the true wave propagation velocity.

This distinction provides a natural explanation for the small discrepancies between the fitted velocities and the theoretical value $v_{\rm SW}+v_{\rm A}$. At the same time, the remarkable linearity of $r_{\rm max}(t)$ and the close correspondence between the measured and expected Alfvénic propagation speeds indicate that the tracked maxima remain closely associated with the propagating Alfvénic structures. The agreement is therefore unlikely to be coincidental and would not be expected for a purely advected feature travelling at $v_{\rm SW}$ alone, nor for a stationary or substantially decelerating structure.

In the present simulation, the plasma $\beta$ of the surrounding environment appears to play a key role in the evolution of the magnitude of the maximum angles of the deflections within a SB event. In the simulations, both the single deflection and the leading deflection of the double-deflection configuration exhibit a three-stage evolution: rapid growth in low-$\beta$ regions, moderate amplification near $\beta \sim 1$, and saturation in high-$\beta$ regions. In contrast, the trailing deflection in the double-deflection case does not follow this trend.
Regarding the dependence of the evolution of the magnitude of the deflection angle to the environmental plasma $\beta$, the differing trends observed between the leading and trailing deflections in the double-deflection scenario can potentially be attributed to the perturbation of the local plasma by the leading deflection, which effectively alters the $\beta$ profile encountered by the trailing deflection. Alternatively, these differences may reflect distinct reconnection conditions at formation—for example, an impulsive primary reconnection event followed by a slower, secondary reconnection. Overall, these results suggest that, in the present configuration, the evolution of the magnitude of the deflection maxima within the SB event is primarily governed by large-scale coronal gradients of Alfvén velocity and plasma $\beta$, rather than by local nonlinear effects alone.

Previous studies \citep[\eg][]{Mallet21, Shi24}, either theoretical or based on simulations, have shown that the nonlinear evolution and stability of SBs are strongly regulated by plasma $\beta$; moreover, SBs become increasingly stable as $\beta$ increases for $\beta$ < 1, while for $\beta$ > 1 they rapidly destabilise as the nature of the compressive fluctuations changes. This behaviour may account for several observed properties of SBs, including their large aspect ratios and their nearly constant occurrence rates in the inner heliosphere.

Thus, the radial profile of the background plasma $\beta$ seems to play a key role in controlling the growth of the magnitude of the maximum of the deflection. In the present simulations, the plasma $\beta$ of the environment increases rapidly with radial distance, raising the possibility that the background profiles may reach high $\beta$ values too quickly. If the plasma $\beta$ were to remain at low values, or close to unity, over a larger radial extent, the maximum deflection angle could continue to grow to greater heights and potentially approach full magnetic field reversals.  Similarly, the use of a bipolar magnetic field profile scaling as $1/r^{2}$ may contribute to this rapid increase of plasma $\beta$. Adopting a different radial dependence of the magnetic field, for instance scaling as $1/r^3$, could maintain plasma $\beta$ values closer to $\beta\sim1$ over a broader range of heights, thereby allowing for a more sustained growth of the deflection amplitude. 

Because the present study focuses on a single simulation setup, the trends identified here should be regarded as physically motivated indications rather than universal properties of switchbacks. Establishing the robustness of the inferred plasma-$\beta$ dependence will require a systematic exploration of alternative background profiles, including those considered in the parametric study of \citetalias{Touresse24}, as well as simulations with more realistic, global solar magnetic configurations, in which the field topology is inherently multipolar. 

Focusing on the formation of full-reversal SBs, and in agreement with the simulation of \citet{Wyper22}, the results in \citetalias{Touresse24} suggest a two-step scenario. First, jet-like events in the low-$\beta$ corona generate torsional Alfvénic waves that propagate along open magnetic field lines. Second, these pre-existing waves are further modified by in-situ processes, such as shear flows \citep[][]{Toth23} and expansion of the surrounding plasma \citep[][]{Squire20, Shoda21}, which can amplify the deflection and produce full magnetic reversals.

Similarly, \citet{Sioulas26} suggests a multi-stage origin for SBs, beginning with coronal fluctuations that grow in amplitude during large-scale expansion below the Alfvén critical point. As these structures are carried outward into the super-Alfvénic solar wind, turbulent evolution modifies their scale-dependent properties, indicating that in-situ formation in the super-Alfvénic regime is not required.

In this context, an important open question is whether full-reversal SBs primarily result from solar-wind expansion itself, or from the persistence of a low-$\beta$ environment over an extended radial range. A natural next step in the numerical modelling is therefore to incorporate an explicit solar-wind expansion term in order to quantify its role in the amplification process. Independently, extending the simulation domain radially and considering atmospheric configurations in which the $\beta = 1$ surface is located at larger heliocentric distances (like the L$\beta$ simulation in \citetalias{Touresse24}) would allow us to assess whether a more extended low-$\beta$ region alone would be sufficient to produce full-reversal SBs.

While the present simulation shows that propagating coronal jets can naturally induce Alfvénic structures with properties consistent with observed SB signatures, they do not exclude alternative formation pathways, nor state their combination to in situ processes \citep[\eg][]{Fisk20, Schwadron21, Squire22, Toth23, Wyper26}. Distinguishing between these different contributions remains an important objective for future observational and numerical studies.

Together, these results provide new insights into the 3D magnetic topology of SB regions and their dependence on local plasma conditions, illustrating how reconnection-driven jet activity in the low corona can generate structures with properties consistent with SBs observed by Parker Solar Probe. They underscore the critical role of plasma variability in shaping the range of SB signatures observed in the solar wind and the intrinsic structural diversity within individual SBs.

\begin{acknowledgements}
The authors thank the referee for all the suggestions and comments, which have constructively helped improve and strengthen the manuscript. We are grateful to Benjamin Lynch and Judy Karpen for the useful discussions. We sincerely thank Valentin Aslanyan for providing access to his ASOT Python routines. JT further thanks Sophie Masson, Guillaume Aulanier, Mohammed Baraka, and Jonas Magrou for valuable discussions. JT acknowledges funding from the Initiative Physique des Infinis of Sorbonne Université. This work was supported by the Action Thématique Soleil-Terre (ATST) of CNRS/INSU PN Astro, co-funded by CNES and CEA, and by financial support from the French national space agency (CNES) through the APR program. This work was also supported by the Agence Nationale de la Recherche (ANR; French National Research Agency) for its financial support of the JET2SB project n°ANR-2025-CE31-7416. CF acknowledges funding from the CEFIPRA Research Project No. 6904-2. PW was supported by a Leverhulme project grant. Essential tests and preparatory simulation runs were made possible thanks to the granted access to the HPC resources of MesoPSL, financed by the Region Île-de-France and the project Equip@Meso (reference ANR-10-EQPX-29-01) of the programme Investissements d’Avenir, supervised by the ANR.  This project was provided with computing HPC and storage resources by GENCI at CINES and IDRIS thanks to the grants   AD010414996, A0160406331 \&  A018040633 on the supercomputer Jean Zay's CLS and ADASTRA's GENOA partitions. Data manipulation was performed using the numpy \citep[][]{VanDerWalt11}, scipy \citep[][]{Virtanen20}, sunpy \citep[][]{sunpy_community2020} and ASOT Python routines. Some figures in this work were produced using the Python package matplotlib \citep[][]{Hunter07}.

\end{acknowledgements}

\bibliographystyle{aa}
\bibliography{JetPropag.bib}

\onecolumn

\begin{appendix}

\section{Diversity of switchbacks profiles sampled by distinct virtual PSP trajectories} \label{app:A}

This appendix presents a set of additional virtual PSP trajectories sampling different regions of the same torsional Alfvénic wavefront.
  
The virtual PSP trajectories are constructed following the procedure described in Section \ref{sec:psp_traj}, using different values for the target location ($r_{1}$, $\theta_{1}$, $\phi_1$) and of the rotation angle around this reference point, $\alpha_{\rm{rot}}$.  The trajectory parameters are listed in Table~\ref{tab:tab_param}.

\begin{table}[h!]
\caption{Parameters defining the virtual PSP trajectories: target location ($r_{1}$, $\theta_{1}$, $\phi_{1}$) and rotation angle $\alpha_{\mathrm{rot}}$.}
\label{tab:tab_param}
\begin{tabular}{|c|c|c|c|c|c|}
\hline

Trajectory  & Associated Figure & $r_{1}$ {[}R$_{\odot}${]} & $\theta_{1}$ {[}$^{\circ}${]} & $\phi_1$ {[}$^{\circ}${]} & $\alpha_{\rm{rot}}$ {[}$^{\circ}${]} \\ \hline
T1 & Figure 2    & $9.864$                    &  $89.58$         & $3.17$                     & $37$                     \\ \hline
T2 & Figure A.1. & $9.864$                  & $89.58$         & $3.17$                      & $143$                                   \\ \hline
T3 & Figure A.2. & $9.864$                  & $92.27$          & $-1.30$                       & $159$                                  \\ \hline
T4 & Figure A.3. & $9.864$                  & $89.84$          & $1.95$                     & $160$                                   \\ \hline
\end{tabular}
\end{table}

\begin{figure*}
\sidecaption
  \includegraphics[width=12cm]{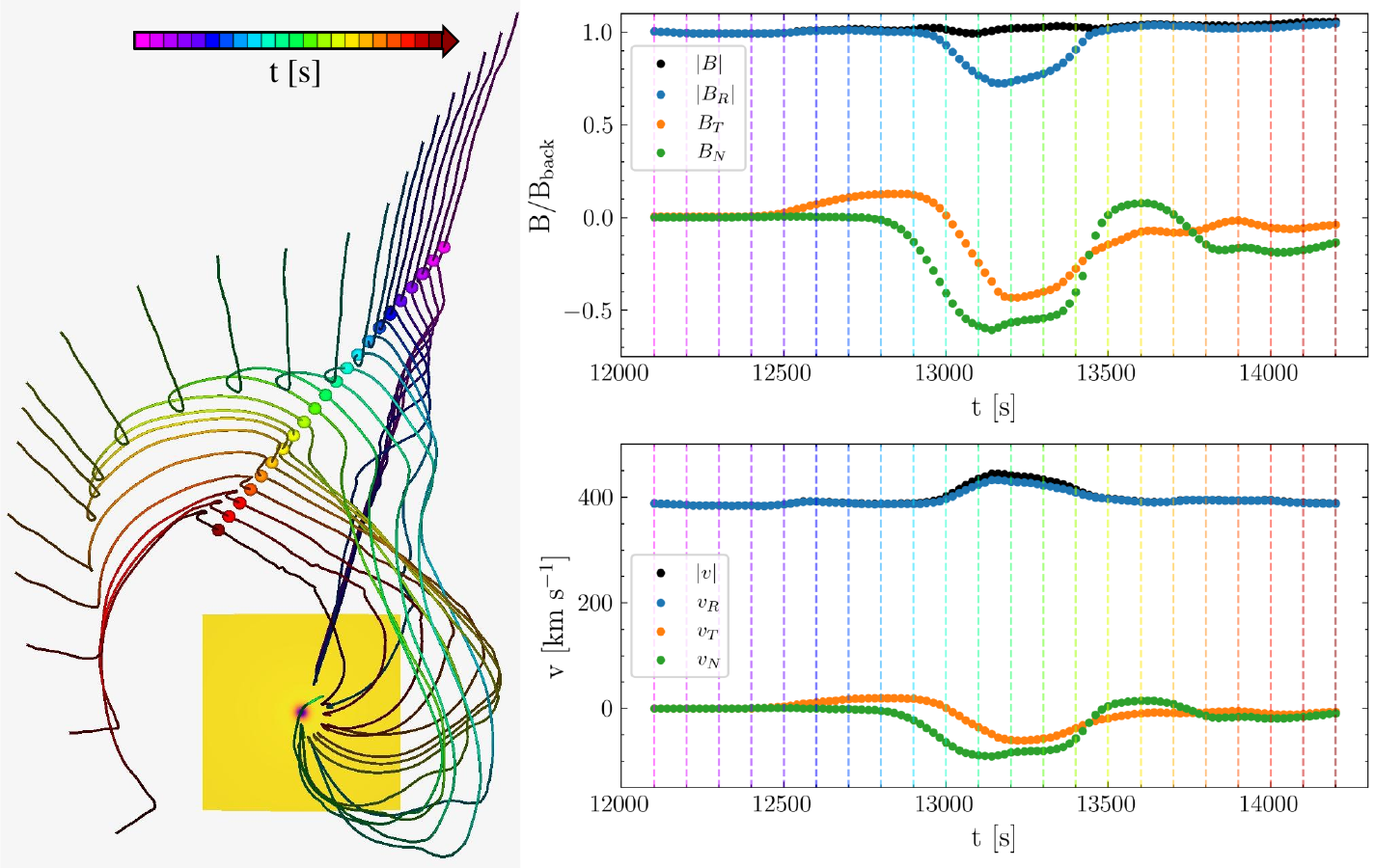}
     \caption{Same as Figure~\ref{fig:synthetic_psp_trajectory}, for the T2 trajectory parameters of Table~\ref{tab:tab_param}. Here, the virtual PSP samples the very same trajectory as in Figure~\ref{fig:synthetic_psp_trajectory}, but with a reverse course compared to the trajectory T1. The associated movie is available online.}
     \label{fig:reverse_trajectory}
\end{figure*}

For each trajectory, we compute the corresponding synthetic in-situ measurements as well as the virtual diachronic magnetic connectivity from the spacecraft (see \Figss{reverse_trajectory}{trajectory_7}; the associated movies show their temporal evolution).
Table~\ref{tab:tab_result} summarises the main properties of the switchback event encountered along each trajectory. 

\begin{table}[h!]
\caption{Main properties of the switchbacks sampled along each virtual PSP trajectory.}
\label{tab:tab_result}
\begin{tabular}{|c|c|c|c|c|c|c|}
\hline
Trajectory & Associated Figure & Duration {[}s{]} & $\delta B_r / B_r$ & $\delta v_r $ {[}km s$^{-1}${]} & $\delta |B| $/ B & $\alpha_{\rm{max}}$ {[}$^{\circ}${]} \\ \hline
T1 & Figure 2   & 900              &     0.33           & 70                           &        0.06          & 46                                   \\ \hline
T2 & Figure A.1.& 540              &       0.28        & 47                           &         0.04         & 45                                   \\ \hline 
T3 & Figure A.2.& 1040             &   0.57                 & 96                            &      0.14            & 64                                   \\ \hline
T4 & Figure A.3.& 500              &      0.36              &          72                       &   0.06               &      51                                \\ \hline
\end{tabular}
\end{table}

These trajectories show the diversity of switchbacks profiles within the same torsional Alfvénic wavefront of a coronal jet. 

 \begin{figure*}
\sidecaption
  \includegraphics[width=12cm]{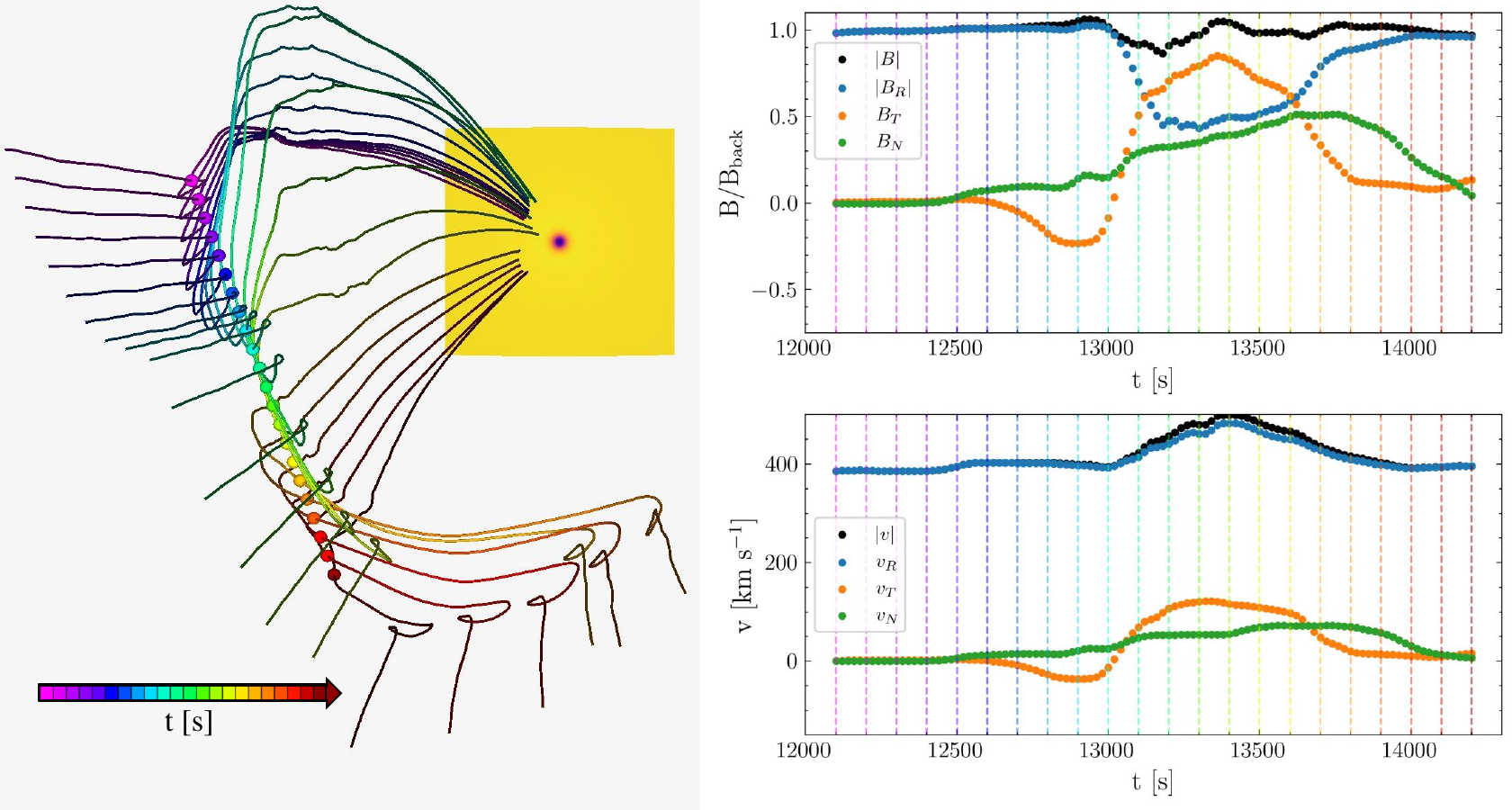}
     \caption{Same as Figure~\ref{fig:synthetic_psp_trajectory}, for the T3 trajectory parameters of Table~\ref{tab:tab_param}. The associated movie is available online.}
     \label{fig:trajectory_2}
\end{figure*}

 \begin{figure*}
\sidecaption
  \includegraphics[width=12cm]{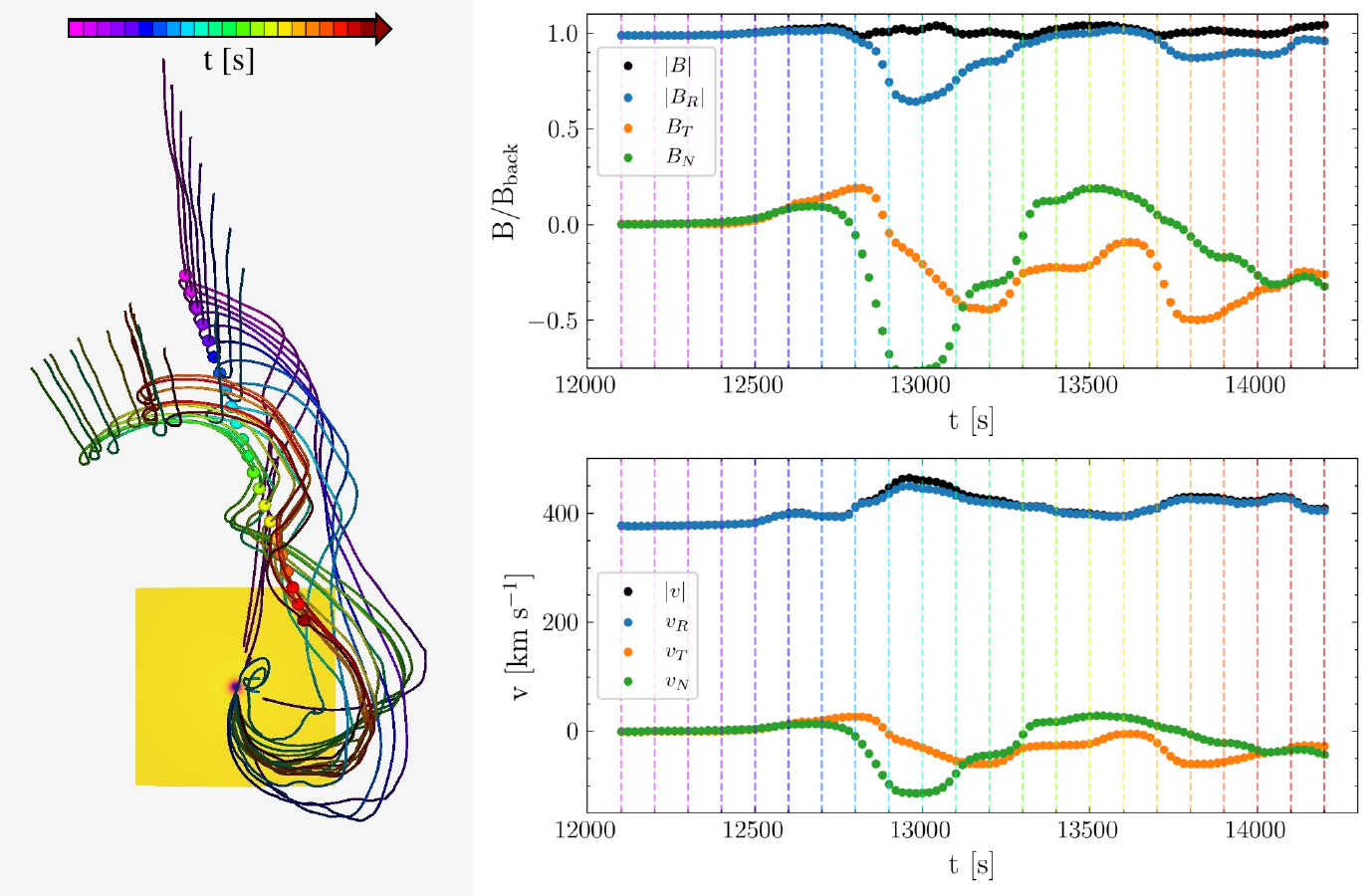}
     \caption{Same as Figure~\ref{fig:synthetic_psp_trajectory}, for the T4 trajectory parameters of Table~\ref{tab:tab_param}. The associated movie is available online.}
     \label{fig:trajectory_7}
\end{figure*}

\end{appendix}

\end{document}